\documentclass[epj]{svjour}
\usepackage{amssymb}
\usepackage{amsmath}
\usepackage{array} % to improve presentation of tables
\usepackage{epsfig}
\usepackage{graphics}
\usepackage{colortbl}
\usepackage{hhline}
\usepackage{multirow}
\newcommand{\SGN}{\mathop{\rm sign}\nolimits}

\begin{document}
\title{Exclusive Double Diffractive Events: general framework and prospects}
\author{R.A.~Ryutin\thanks{\emph{e-mail:} Roman.Rioutine@cern.ch}\inst{1}
}                     % Do not remove
\institute{{\small Institute for High Energy Physics},{\small{\it 142 281}, Protvino, Russia}}
\abstract{
We consider the general theoretical framework to study
exclusive double diffractive events (EDDE). It is
a powerful tool to explore the picture of the $pp$ 
interaction. Basic kinematical and dynamical properties of the process, and
also normalization of parameters via standard processes like
the exclusive vector meson production (EVMP), are considered in detail.
As an example, calculations of the cross-sections
in the model with three pomerons
for the process $p+p\to p+M+p$ are presented for Tevatron and LHC energies. 
 % end of PACS codes
} %end of abstract
\authorrunning
\titlerunning
\maketitle
%

%----------------Introduction---------------------------------------------

\section{Itroduction}

  Impressive progress of the LHC ex\-pe\-ri\-ments stimulates new investigations in different areas of high energy physics. Latest weighty arguments
 in favor of the existence of the Higgs boson give us the basis for futher ex\-pe\-ri\-ments, since we have to define exactly the nature
 of this particle. It is a problem of today to have a clear instrument to determine its quantum numbers and couplings to other particles. Besides 
 this task concerning Higgs boson, we have to continue to study other fundamental objects of high energy physics like jets and particles
 produced in different processes. 
 
  If we consider huge number of high energy processes which have been studied for a long time, we will find the one which can serve as a 
 clear source of information about high-energy dynamics. It is the exclusive double diffractive event (EDDE), i.e. the process of the type
 $h+h\to h^*+M+h^*$, where $h\to h^*$ scattering is quasi-diffractive, $M$ is the centrally produced 
 particle or system of particles and ``$+$'' means large rapidity gap (LRG). If one takes $M$ as a single particle produced, this is the first 
 ``genuinely'' inelastic process which not only retains a lot of features of elastic scattering (diffractive patterns), but also shows clearly how the initial 
 energy is being transformed into the secondary particles. General properties of such amplitudes were considered in Ref.~\cite{old1}-\cite{old3}. Theoretical 
 consideration of these processes on the basis of Regge theory goes back to papers~\cite{oldmodels1}-\cite{oldmodels5}. The recent interest is related 
 to possibly good signals of centrally produced Higgs 
 bosons, heavy quarkonia, glueballs, jets, gauge 
 bosons, system of hadrons~\cite{newmodels1}-\cite{newmodels10}.
 
 Experimental study has begun since 1970's~\cite{oldexp1}. As Pomerons are the driving force of the processes in question at high energies 
 it is naturally to expect that glueball production will be favorable, if one believes that Pomerons are mostly gluonic objects. Central glueball production was suggested as possible origin of the total cross section rise in Ref.~\cite{old3}. One of the early proposal for ex\-pe\-ri\-mental investigations of centrally produced glueballs in EDDE was made in~\cite{oldexp2}. As to the most recent ex\-pe\-ri\-mental studies one has to mention the series of results from the ex\-pe\-ri\-ment WA102~\cite{WA102}-\cite{WA102d}. With energy rise
 we can observe central systems with higher masses~\cite{CDFreview}, like di-jet~\cite{CDFjj}, di-gamma~\cite{CDF2gamA},\cite{CDF2gamB},\cite{LHC2gamCMS}, heavy quarkonia~\cite{CDFhq}, di-hadron~(\cite{KMRhh} and references therein).
  
  EDDE gives us unique ex\-pe\-ri\-mental possibilities for particle searches
  and investigations of diffraction proper. This is due to several advantages of the 
  process: 
  \begin{itemize}
  \item clear signature of the process: central system separated from two finally detected protons by 
  LRGs (see, for example, theoretical work~\cite{LRGs}, the ex\-pe\-ri\-mental one~\cite{LRGs2} and references therein); 
  \item possibility to use the ``missing mass method''~\cite{MMM} that improves the mass 
  resolution; 
  \item strong suppression of the background due to the\linebreak $J_z=0$ selection rule~\cite{Jz0Pumplin}-\cite{Jz0Khoze2} for basic processes like, for example, Higgs boson production; 
  \item spin-parity analysis of the central system can be 
  done to determine quantum numbers of the central 
  particle~\cite{spin_parity_analyser1}-\cite{spin_parity_analyser4}; 
  \item interesting 
  measurements concerning the interplay between ``soft'' and ``hard'' scales are possible: we can obtain basic features of the interaction
  region (size and shape) from distributions in the scattering angle (diffractive patterns)~\cite{diff_patterns}.
  \end{itemize}
%=== first LHC results ===   
  There are several proposals to realize the above properties at LHC~\cite{CMS_TOTEM}-\cite{LHCLRGs}. Due to the 
  complicated picture of the interaction at high luminocities (a lot of ``pile-up'' events) at the moment we only have the 
  possibility to select EDDE without detection of final protons~\cite{LHCnopptag},\cite{LHC2gamCMS}. The criterium of LRGs is not sufficient for our investigations, since we loose
  basic advantages of the exclusive process. For the ex\-pe\-ri\-ment we need special low luminocity runs. In fact, latest LHC ex\-pe\-ri\-ments show that the definition 
  of diffraction is rather complicated
  subject~\cite{LHCdiffdefine}, which needs futher investigations.
  
%=== other future ex\-pe\-ri\-ments
A new physics program based on po\-la\-ri\-zed pro\-ton be\-ams and tagging of forward protons has been launched also at STAR/RHIC~\cite{RHIC1}-\cite{RHIC4}. These 
ex\-pe\-ri\-ments can drastically improve our understanding of diffractive mechanisms and suppress uncertainties that can reach sometimes an order of magnitude.
  
%===  HERA EVMP, {WA102, CDF} EDDE st.candles - normalization ===
  There are many theo\-re\-ti\-cal groups now that work in this area~\cite{newmodels1}-\cite{newmodels10}. All these models need to
  obtain va\-lu\-es of their parameters to make predictions for the LHC energies. For this purpose we can use so called ``standard candle'' processes, i.e. events
  which have the same theo\-re\-ti\-cal ingredients for the calculations. Usually authors use the following processes:
  
  for high central masses ($M\gg 1$~GeV, perturbative mechanism of Pomeron-Pomeron fusion dominates)
  \begin{itemize}
  \item $\gamma^*+p\to V+p$ (Exclusive Vector Meson Production, EVMP), $m_V\gg 1$~GeV~\cite{HERAevmp1}-\cite{HERAevmp2a};
  \item $p+p\to p+M+p$, $M=jj$~\cite{CDFjj}, $M=\gamma\gamma$~\cite{CDF2gamA},\cite{CDF2gamB},\cite{LHC2gamCMS}, $M=\{Q\bar{Q}\}$ (heavy quarkonia, $\chi_{c,b}$)~\cite{CDFhq}, $M=hh$ (dihadron system)~\cite{KMRhh};
  \end{itemize}
  
  for low central masses ($M\sim 1$~GeV, nonperturbative Pomeron-Pomeron fusion)
  \begin{itemize}
    \item $\gamma^*+p\to V+p$ (EVMP), $m_V\sim 1$~GeV~\cite{HERAevmp3}-\cite{HERAevmp4};
    \item $p+p\to p+M+p$, $M=\{q\bar{q}\}$ (light meson) or ``glueball''~\cite{WA102}-\cite{WA102d},  $M=hh$ (dihadron system)~\cite{KMRhh};
  \end{itemize}
  In the low-mass case the mechanism of Pomeron-Pomeron fusion is considerably nonperturbative. It will be taken into account in the calculations.
  
%=== Plan of the article ===
   
   In this paper we try to consider a general model-in\-de\-pen\-dent framework for EDDE. Our model for diffractive processes is taken as an example.
  
\section{Exclusive vector meson photoproduction.}
\label{sec:evmpGEN}

 In this chapter we consider EVMP, i.e. exclusive photoproduction of $V=Q\bar{Q}_{1S}$ states, as the first ``standard candle'' for the 
 EDDE. It was considered in~\cite{3P:EVMP2005} in the framework of the three Pomeron model~\cite{3P:v1}. Here we present a more general
 situation and correct formulae for parameters and integrals. Basic ingredients of the theoretical framework are presented in the Fig.~\ref{fig:EVMPall} and in 
 subsections~\ref{sec:evmpKIN}-\ref{sec:evmp3P}. Below we consider perturbative mechanism of EVMP, i.e. the mass of a vector meson is much greater than $1$~GeV in this 
 case. Nonperturbative mechanisms (see Fig.~\ref{fig:EVMPother}) will be presented in futher works. Following the scheme depicted in the Fig.~\ref{fig:EVMPall}
 we can write the amplitude of the process as a convolution of the diffractive gluon-proton amplitude $T$ and perturbative amplitude $A$, which is considered
 in an appropriate approach.

\subsection{Kinematics.}
\label{sec:evmpKIN}
\begin{figure}  
 % \hspace*{2cm}
  \includegraphics[width=0.49\textwidth]{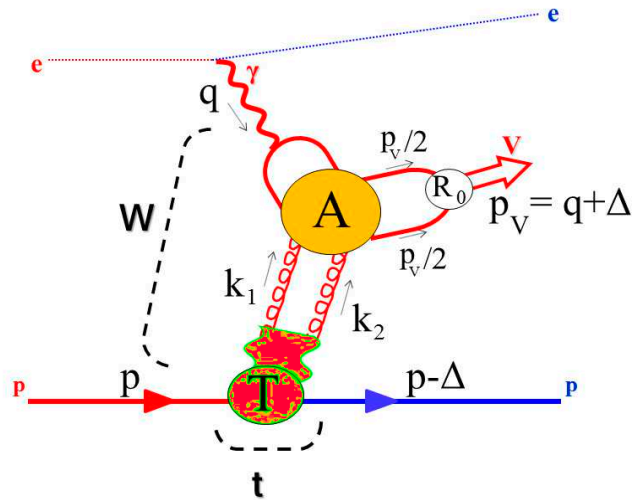} 
  \caption{\label{fig:EVMPall}Exclusive vector meson photoproduction in the NRQCD framework. $W$ is the $\gamma p$ collision energy. $t=-\Delta^2$ is the transferred momentum squared. Directions of momenta are shown by arrows. Vertexes $A$, $T$ and $R_0$ are defined in the text.}
\end{figure}

First, let us introduce the notations for four-vectors depicted in the Fig.~\ref{fig:EVMPall}. For any four-vector
we use two following representations,
\begin{eqnarray}
&&\mbox{usual:}\; v=\left( v_0,v_3; \vec{v}_t\right)\;\nonumber\\
&&\label{v_represent}\mbox{and ``light-cone'':}\; v=\left\{ v_{+},v_{-}; \vec{v}_t\right\}.
\end{eqnarray}

For external vectors we have:
\begin{eqnarray}
\label{p_exact}&& p=\left( \frac{W^2+Q^2+m^2}{2W}, \frac{\lambda^{1/2}\left( W^2,m^2,-Q^2\right)}{2W}; \vec{0}\right),\\
\label{q_exact}&& q=\left( \frac{W^2-Q^2-m^2}{2W}, -\frac{\lambda^{1/2}\left( W^2,m^2,-Q^2\right)}{2W}; \vec{0}\right),\\
&&\Delta_3=\frac{W^2(m_V^2+Q^2-3t)+(m^2+Q^2)(m_V^2+Q^2+t)}{2\lambda^{1/2}\left( W^2,m^2,-Q^2\right) W},\nonumber\\
\label{delta_exact}&& \Delta=\left( \frac{m_V^2+Q^2+t}{2W},\; \Delta_3;\; \vec{\Delta} \right),
\end{eqnarray}
where
\begin{eqnarray}
\label{lambda_fun}&&\lambda(x,y,z)=x^2+y^2+z^2-2xy-2xz-2yz,\\
\label{momenta_squared_EVMP}&& s=W^2,\; q^2=-Q^2,\; p^2=m^2,\; \Delta^2=t,\; p_V^2=m_V^2.
\end{eqnarray}
Let us introduce some basic variables:
\begin{eqnarray}
&&k_1=\kappa+\frac{\Delta}{2},\;
k_2=-\kappa+\frac{\Delta}{2},\;\nonumber\\
&&p'=p-\frac{m^2}{\tilde{s}}q,\;
q'=q+\frac{Q^2}{\tilde{s}}p,\; p_V=q+\Delta,\;\nonumber\\
&& q'^2=p'^2=0,\; 
\tilde{s}=pq+\sqrt{(pq)^2+Q^2m^2},\nonumber\\
&&\kappa=\frac{z_v}{2}\left(y_{+} p'+y_{-} q'\right)+\kappa_{\perp},\;\nonumber\\
&& \Delta=z_v^2\left(\left[ 1+y_Q+y_{\Delta}\right]p'-y_{\Delta}q'\right)+\Delta_{\perp},\;\nonumber\\
&& z_v=\frac{m_V}{W},\; y=-\frac{4\kappa^2}{m_V^2},\;
y_Q=\frac{Q^2}{m_V^2},\;\nonumber\\
&& t\simeq\Delta_{\perp}^2=-\vec{\Delta}^2,\;
y_{\Delta}=\frac{\vec{\Delta}^2}{m_V^2},\; y_0=\frac{4s_0}{m_V^2},\;\nonumber\\
&&\kappa_{\perp}^2=-\vec{\kappa}^2=
-\frac{m_V^2}{4}\left( y+4y_{+}y_{-}\frac{2p'q'}{W^2}\right)\nonumber\\
\label{nota}&&\simeq 
-\frac{m_V^2}{4}\left( y+4y_{+}y_{-}\right),
\end{eqnarray}
$s_0=1\;\mbox{GeV}^2$.

Photon and vector meson polarization vectors in the general case ($Q\neq0$) can be represented as follows:
\begin{eqnarray}
&&{\epsilon_{\gamma}}_{\perp}q={\epsilon_{\gamma}}_0q=0\;,\;{\epsilon_{\gamma}}_{\perp}^2=-{\epsilon_{\gamma}}_0^2=-1\;,\nonumber\\
&& {\epsilon_{\gamma}}_0=\frac{1}{Q}(q'+z_v^2 y_Q p')\;,\nonumber\\
&&{\epsilon_V}_{\perp}p_V={\epsilon_V}_{\parallel}p_V=0\;,\nonumber\\
&&\;{\epsilon_V}_{\perp}=v_{\perp}+\frac{2(\vec{v}\vec{\Delta})}{s}(p'-q')\;,\;v_{\perp}^2=-\vec{v}^2\;,\;\nonumber\\
\label{es}&&{\epsilon_V}_{\parallel}=\frac{1}{m_V}(q'-z_v^2(1-y_{\Delta})p'+\Delta_{\perp})
\end{eqnarray}

For high-energy photoproduction ($Q\to 0$) we have 
\begin{eqnarray}
&& Q\ll m,\;m_V,\;\sqrt{-t}\ll W,\nonumber\\
&& z_m=\frac{m}{W},\; z_Q=\frac{Q}{W},\; z_t=\frac{\sqrt{-t}}{W},\nonumber\\
&& k_1\simeq\left\{ 
\frac{W}{\sqrt{2}}z_v\left( y_{+}+\frac{z_v}{2}\right),\;
\frac{W}{\sqrt{2}}z_vy_{-};\; \vec{\kappa}
\right\},\;\nonumber\\
&& k_2\simeq\left\{ 
-\frac{W}{\sqrt{2}}z_v\left( y_{+}-\frac{z_v}{2}\right),\;
-\frac{W}{\sqrt{2}}z_vy_{-};\; -\vec{\kappa}
\right\},\;\nonumber\\
&& p_V\simeq\left\{
\frac{W}{\sqrt{2}}(z_v^2+z_t^2), \frac{W}{\sqrt{2}};\; \sqrt{-t}\frac{\vec{\Delta}}{|\vec{\Delta}|}
\right\},\nonumber\\
&& \label{eq:kinq0}p\simeq p'\simeq\left\{ \frac{W}{\sqrt{2}},\; 0;\; \vec{0} \right\},\;
q\simeq q'\simeq\left\{ 0,\; \frac{W}{\sqrt{2}},\; \vec{0} \right\}.
\end{eqnarray}
It is convenient to introduce two transverse polarization vectors
\begin{eqnarray}
&&\epsilon_1=(0,\;0;\; (1,\;0)),\;\epsilon_2=(0,\;0;\; (0,\;1)),\;\nonumber\\
&& \epsilon_i\perp p,\;q,\;p_V,\; \epsilon_i^2=-1,\;\nonumber\\
\label{es12}&& \epsilon_ik_1=-\epsilon_ik_2=\epsilon_i\kappa=-\left\{ 
\begin{array}{c}
\kappa_t\cos\phi\\
\kappa_t\sin\phi
\end{array}
\right.
\end{eqnarray}

\subsection{Soft and hard amplitudes.}
\label{sec:evmpDYN}
\subsubsection{NRQCD framework for the hard part of the amplitude.}
 
\begin{figure}  
  \includegraphics[width=0.49\textwidth]{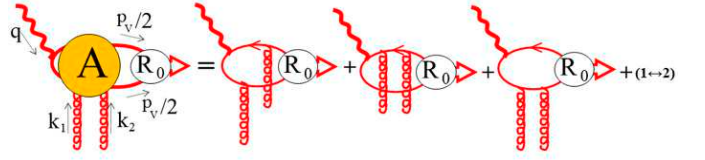} 
  \caption{\label{fig:EVMPhard}Exclusive vector meson photoproduction in the NRQCD framework. Hard amplitude.}
\end{figure}

 The amplitude $A$ of the process $\gamma(q)+g(k_1)\to V(p_V)+g(-k_2)$ (Fig.~\ref{fig:EVMPhard}) is calculated in the nonrelativistic bound state 
approximation (see~\cite{QQth1}-\cite{QQth3} and ref. therein):

\begin{eqnarray}
\tilde{O}^{\alpha\beta}_{\epsilon_{\gamma},\epsilon_V}&\!=&\!
Sp\left[\hat{\cal{O}}^{\alpha\beta}(\hat{p}_V-m_V)\hat{\epsilon}_V\right], \;
{\cal{K}}_V=\frac{8R_{V,0}\pi e_Q\sqrt{\alpha_e} \alpha_s}{\sqrt{3 m_V}},\nonumber\\
A^{\mathrm{ab},\;\alpha\beta}&\!=&\!\frac{R_{V,0}}{4\sqrt{\pi m_V}}\times e_Q\sqrt{4\pi\alpha_e}\times 4\pi\alpha_s\times 
\left[ \mathrm{t}^{\mathrm{a}}_{\mathrm{ij}}\mathrm{t}^{\mathrm{b}}_{\mathrm{jk}} \frac{\delta^{\mathrm{ik}}}{\sqrt{3}}\right] \tilde{O}^{\alpha\beta}_{\epsilon_{\gamma},\epsilon_V}\nonumber\\
&\!=&\!
{\cal{K}}_V \frac{\delta^{\mathrm{ab}}}{8} 
 \tilde{O}^{\alpha\beta}_{\epsilon_{\gamma},\epsilon_V},
\end{eqnarray}
where $e_Q$ is the charge of heavy quark $Q$, $R_{V,0}$ is the absolute value of the
vector meson radial wave function at the origin, $\left[ \mathrm{t}^{\mathrm{a}}_{\mathrm{ij}}\mathrm{t}^{\mathrm{b}}_{\mathrm{jk}} \frac{\delta^{\mathrm{ik}}}{\sqrt{3}}\right]=\frac{\delta^{\mathrm{ab}}}{2\sqrt{3}}$ due to SU(3) group rules and
\begin{eqnarray}
\hat{\cal{O}}^{\alpha\beta}=\!\!\!&&\left\{ 
\frac{\gamma^{\alpha}\left( -\hat{k}_1+\frac{\hat{p}_V+m_V}{2}\right)\hat{\epsilon}_{\gamma}\left( \hat{k}_2+\frac{-\hat{p}_V+m_V}{2}\right)\gamma^{\beta}}{\left( -p_Vk_1+k_1^2+\mathrm{i}\;0\right)\left( -p_Vk_2+k_2^2+\mathrm{i}\;0\right)}
\right.\nonumber\\
&& \left. 
+\frac{\epsilon_{\gamma}\left( \hat{k}_1+\hat{k}_2+\frac{-\hat{p}_V+m_V}{2}\right)\gamma^{\alpha}\left( \hat{k}_2+\frac{-\hat{p}_V+m_V}{2}\right)\gamma^{\beta}}{\left( -p_V(k_1+k_2)+(k_1+k_2)^2+\mathrm{i}\;0\right)\left( -p_Vk_2+k_2^2+\mathrm{i}\;0\right)}
\right.
\nonumber\\
&& \left. 
+\frac{\gamma^{\alpha}\left( -\hat{k}_1+\frac{\hat{p}_V+m_V}{2}\right)\gamma^{\beta}\left( -\hat{k}_1-\hat{k}_2+\frac{\hat{p}_V+m_V}{2}\right)\hat{\epsilon}_{\gamma}}{\left( -p_Vk_1+k_1^2+\mathrm{i}\;0\right)\left( -p_V(k_1+k_2)+(k_1+k_2)^2+\mathrm{i}\;0\right)}
\right\}\nonumber\\
&& + (1\leftrightarrow 2).
\end{eqnarray}
 After calculations we have
 \begin{eqnarray}
  &&\tilde{O}^{\alpha\beta}_{\epsilon_{\gamma},\epsilon_V}=-16m_V\left\{ \phantom{p_V^{\beta}}\!\!\!\!\!\!\!
  k_1k_2\right.\nonumber\\
  &&\left.\times\left[
  p_V^{\alpha}\left( \epsilon_V^{\beta}k_2\epsilon_{\gamma}-\epsilon_{\gamma}^{\beta}k_2\epsilon_V \right)
  +p_V^{\beta}\left( \epsilon_V^{\alpha}k_1\epsilon_{\gamma}-\epsilon_{\gamma}^{\alpha}k_1\epsilon_V \right)
 \right]
 \right.
  \nonumber\\
  && \left.
  +\left(
  \epsilon_V^{\alpha}\epsilon_{\gamma}^{\beta}k_1k_2
 +g^{\alpha\beta}k_1\epsilon_V k_2\epsilon_{\gamma}
  \right)
  \left(
  k_1^2-k_1p_V+k_1k_2
  \right)
  \right.
  \nonumber\\
  && \left.
   +\left(
   \epsilon_V^{\beta}\epsilon_{\gamma}^{\alpha}k_1k_2
  +g^{\alpha\beta}k_2\epsilon_V k_1\epsilon_{\gamma}
   \right)
   \left(
   k_2^2-k_2p_V+k_1k_2
   \right)-\epsilon_V\epsilon_{\gamma}
   \right.
  \nonumber\\
  && \left.
  \times\left[
  g^{\alpha\beta}
  \left(
  k_1^2-k_1p_V+k_1k_2
  \right)
  \left(
  k_2^2-k_2p_V+k_1k_2
   \right)
  +p_V^{\alpha}p_V^{\beta}k_1k_2
  \right]
  \right\}\nonumber\\
  &&\times\left[\left( -p_Vk_1+k_1^2+\mathrm{i}\;0\right)\left( -p_Vk_2+k_2^2+\mathrm{i}\;0\right) \right.\nonumber\\
  &&\left.
 \left( -p_V(k_1+k_2)+(k_1+k_2)^2+\mathrm{i}\;0\right)\right]^{-1}. 
 \end{eqnarray}

\subsubsection{Diffractive part of the amplitude.}

Diffractive gluon-proton amplitude can be written as 
\begin{eqnarray}
 &&T^{\mathrm{ab},\;\alpha\beta}\equiv 
 T^{\mathrm{ab},\;\alpha\beta}(p,k_1,k_2)\nonumber\\
 \label{eq:T}&&=
 \delta^{\mathrm{ab}} 
 \tilde{G}^{\alpha\beta}(p,k_1,k_2) T^D ((p-k_1)^2,t,k_1k_2),
\end{eqnarray}
where
\begin{eqnarray}
&&\tilde{G}^{\alpha\beta}= G^{\alpha\beta}-
\frac{P_1^{\alpha}P_2^{\beta}}{P_1P_2},\;\nonumber\\
&& G^{\alpha\beta}=g^{\alpha\beta}-\frac{k_2^{\alpha}k_1^{\beta}}{k_1k_2},\;
P_{1,2}=p-k_{2,1}\frac{pk_{1,2}}{k_1k_2}.
\end{eqnarray}
Strictly speaking, in the general case we have to write more than one transverse tensor structures, but
in most cases these 
structures satisfy relations like the Callan-Gross one~\cite{CallanGross1},\cite{CallanGross2}. Here 
we can neglect terms of the order $o(t)$ and $o(z_v^2)$. Further 
we will take into account relations
\begin{equation}
\label{eq:Tgaugeinvar}k_{1,\;\alpha}T^{\mathrm{ab},\;\alpha\beta}=0,\; k_{2,\;\beta}T^{\mathrm{ab},\;\alpha\beta}=0,
\end{equation}
that allow us to replace the tensor part of gluon propagators 
by metric tensors.

Let us denote
\begin{equation}
{\cal{T}}_{\epsilon_{\gamma},\epsilon_V}=\tilde{O}^{\alpha\beta}_{\epsilon_{\gamma},\epsilon_V}
d_{\alpha\alpha'}d_{\beta\beta'}\tilde{G}^{\alpha'\beta'},
\end{equation}
where
\begin{eqnarray}
&&d_{\alpha\alpha'}=-g^{\alpha\alpha'}
+\frac{1}{(k_1k_2)^2-k_1^2k_2^2}\nonumber\\
&&\times\left[ 
k_1k_2(k_{1,\;\alpha}k_{2,\;\alpha'}+k_{1,\;\alpha'}k_{2,\;\alpha})\right.\nonumber\\
&&\left.
-k_1^2k_{2,\;\alpha}k_{2,\;\alpha'}-k_2^2k_{1,\;\alpha}k_{1,\;\alpha'}
\right]
\end{eqnarray}
is the tensor part of the gluon propagator in the axial gauge with the axial vecor $n=k_1+k_2=\Delta$ and
$$
\vec{\Delta}\to 0 \Longrightarrow t\to 0 \Longrightarrow n^2\to 0.
$$
We can obtain the main (diagonal) contribution for further calculations when we set $\epsilon_{\gamma}=\epsilon_V=\epsilon_{i}$ ($i=1,2$)
\begin{eqnarray}
&&{\cal{T}}_{(i,i)}= \frac{8f_c(y,y_{+},y_{-})}{m_V\left( y_{-}^2+z_m^2 y \right)}\nonumber\\
&&\times\frac{1}{
\left( y_{+}-\frac{z_v}{2}(1+y)+\mathrm{i}\;0 \right)
\left( y_{+}+\frac{z_v}{2}(1+y)-\mathrm{i}\;0 \right)
},\\
&&  f_c(y,y_{+},y_{-})=
y^2-2z_v^2y_{-}^2(1+(1+2 c^2)y)\nonumber\\
\label{eq:fc}&&\hspace*{2cm}+4y_{-}y_{+}(y-4c^2z_v^2y_{-}^2)+8y_{-}^2y_{+}^2.
\end{eqnarray}
In the above equation~(\ref{eq:fc}) $c$ is the corresponding trigonometric function $\cos\phi$ ($i=1$) or $\sin\phi$ ($i=2$). In the next subsection 
we will have to take the integral over $\phi$, which is equivalent to the replacement $c^2\to 1/2$ that
is why it is convenient to introduce the following function
\begin{eqnarray}
&&\hspace*{-0.9cm}f(y,y_{+},y_{-})=\int\limits_{0}^{\pi}\frac{d\phi}{\pi} f_c(y,y_{+},y_{-})\nonumber\\
&&\hspace*{-0.9cm}=y^2-2z_v^2y_{-}^2(1+2y)+4y_{-}y_{+}(y-2z_v^2y_{-}^2)+8y_{-}^2y_{+}^2.
\end{eqnarray}
 
\subsubsection{Convolution and integration. Extraction of parameters for the diffractive amplitude.}

After all convolutions the diagonal element of the amplitude for the process $\gamma+p\to V+p$ looks as follows
\begin{eqnarray}
{\cal M}_{i,i}&=&\int\frac{d^4\kappa}{(2\pi)^4}\frac{{\cal{K}}_V {\cal{T}}_{(i,i)} T^D((p-k_1)^2)}{(k_1^2+\mathrm{i}\;0)(k_2^2+\mathrm{i}\;0)}\nonumber\\
&=& \int dy\;dy_{+} dy_{-} \int\limits_{0}^{\pi}\frac{d\phi}{\pi}
\frac{\pi m_V^4}{8}\frac{{\cal{K}}_V {\cal{T}}_{(i,i)} T^D(z_v y_{-} s)}{(2\pi)^4}\nonumber\\
&\times&\left[ 
\frac{4}{m_V^4z_v^2\left(y_{-}^2-\left( \tilde{y}_{-}-\mathrm{i}\;0 \right)^2\right)}
\right]
\nonumber\\
\label{eq:amplitude}&=& \overline{{\cal{K}}}_V {\cal{I}}_V \\
{\cal{I}}_V&=& \frac{1}{8}\int dy\;dy_{+} dy_{-} \frac{f(y,y_{+},y_{-})}{z_v^2\left( y_{-}^2+z_m^2 y \right)}\nonumber\\
\label{eq:genIv1}&\times&\frac{T^D(z_v y_{-} s)}{\left(y_{-}^2-\left( \tilde{y}_{-}-\mathrm{i}\;0 \right)^2\right) \left( y_{+}^2-\left(\tilde{y}_{+}-\mathrm{i}\;0\right)^2 \right)},\\
\overline{{\cal{K}}}_V&=&\frac{16R_{V,0}e_Q\sqrt{\alpha_e} \alpha_s}{\sqrt{3}m_V^{3/2}\pi^2},\;\\
\tilde{y}_{-}&=&\left|\frac{y}{2z_v}\right|,\; \tilde{y}_{+}=\left|\frac{z_v}{2}(1+y)\right|.
\end{eqnarray}
Integral~(\ref{eq:genIv1}) can be represented in the following form
\begin{eqnarray}
\label{eq:genIv2}{\cal{I}}_V&\simeq & \frac{1}{8}\int\limits_0^1 dy 
\left\{ 
\tilde{f}_{+}\hat{{\cal{I}}}_{+}+\tilde{f}_{-}\hat{{\cal{I}}}_{-}
\right\},\\
\tilde{f}_{\tau}&= &f(\tilde{y}_{+},\tilde{y}_{-},y)=\frac{y^2}{2}
\left[ 
2+y^2+\tau |1+y|(2-y)
\right],\;\nonumber\\
\tau&=&\SGN\left( y_{+}y_{-}y\right),\; \tilde{f}_{\pm}\equiv\tilde{f}_{\pm 1},\\
\hat{{\cal{I}}}_{\pm}&\simeq & \int\limits_0^{\infty} 
\frac{dy_{+}}{\left( y_{+}^2-  \left( \tilde{y}_{+}-\mathrm{i}\;0\right)^2  \right)}
\int\limits_0^{\infty}
\frac{dy_{-}}{\left( y_{-}^2-  \left( \tilde{y}_{-}-\mathrm{i}\;0\right)^2  \right)}\nonumber\\
&\label{eq:hatgenIv2}\times & \frac{T^D(z_vy_{-} s)+T^D(-z_v y_{-} s)}{z_v^2 y_{-}^2} \theta\left( y\pm 4y_{+}y_{-}\right),
\end{eqnarray}
where $1/z_v$ is replaced by infinities in the upper limits, and $z_m$ is set to zero in the denominator 
$(y_{-}^2+z_m^2 y)$. These replacements do not change much the final result, but significantly
simplify further calculations. Contribution of residues in~(\ref{eq:hatgenIv2}) is dominant, if 
we take into account only the imaginary part of the amplitude $T^D$ (which is 
rather good approximation in most interesting cases), and
we can write in this case
\begin{equation}
\label{eq:genIv2ImTD}{\cal{I}}_V\simeq -\frac{\pi^2}{2}\int\limits_0^1 \frac{dy}{y(1+y)}(y+4)T^D(\frac{y\; s}{2}).
\end{equation}
As to the denominator, there are simple arguments to set $z_m$ to zero. Since $T^D(0)=0$, and the contribution from 
the region $s\;y/2>m^2$ (i.e. $y>2z_m^2$) is dominant, we have
\begin{eqnarray}
&&y_-^2+y z_m^2\simeq y\left( \frac{y}{4z_v^2}+z_m^2\right),\nonumber\\
&&\label{eq:denomexpl} \frac{y}{4z_v^2}>\frac{z_m^2}{2z_v^2}\gg z_m^2.
\end{eqnarray}

Now we can extract parameters for the amplitude $T^D$ of the process $g+p\to g+p$. If we use the model of vector 
dominance, the amplitude of the photoproduction looks as
\begin{equation}
\label{eq:VDMamplitude} {\cal M}_{i,i}^{\gamma+p\to V+p}=\frac{3\pi^2}{8\alpha_s(m_V^2)}\overline{\cal{K}}_V {\cal M}_{i,i}^{V^*+p\to V+p}. 
\end{equation}
We can fix parameters of a model for meson-proton scattering by fitting the data on photoproduction. Then we can 
use expression~(\ref{eq:amplitude}) and compare it with~(\ref{eq:VDMamplitude}). Finally
we can obtain the amplitude of gluon-proton scattering from the equality
\begin{equation}
\label{eq:masterEVMP}{\cal{I}}_V=\frac{3\pi^2}{8\alpha_s(m_V^2)} {\cal M}_{i,i}^{V^*+p\to V+p}.
\end{equation}

In the above expressions we do not take into account the dependence of ${\cal{I}}_V$ on $t$, since it is rather weak. It comes from
the diffractive amplitude $T^D\equiv T^D(s,t)$. It means that~(\ref{eq:amplitude}),(\ref{eq:genIv1}),(\ref{eq:hatgenIv2}), (\ref{eq:genIv2ImTD})  and~(\ref{eq:masterEVMP}) are taken at any fixed value of $t$. If we use the data on total cross-sections we
have to use the following equality:
\begin{eqnarray}
&&\int\limits_{t_{\mathrm{min}}}^{t_{\mathrm{max}}}dt\left| {\cal{I}}_V(s,t) \right|^2\nonumber\\
\label{eq:masterEVMPt}&=&
\left( \frac{3\pi^2}{8\alpha_s(m_V^2)}\right)^2 \int\limits_{t_{\mathrm{min}}}^{t_{\mathrm{max}}}dt\left| {\cal M}_{i,i}^{V^*+p\to V+p}(s,t)\right|^2.
\end{eqnarray}

\subsection{3-Pomeron model as an example for calculations.}
\label{sec:evmp3P}
Let us stress that in this paper calculations of integrals do not depend on the form of gluon-proton scattering 
amplitude $T^D$. In principle, we can use 
any model it. For example, in the case
of the ordinary unintegrated gluon distribution
\begin{equation}
\label{eq:TDandfg}T^D (\frac{y\; s}{2})\simeq T^D(\frac{\vec{\kappa}^2}{x})\Longrightarrow f_g(x,\vec{\kappa}^2,\mu^2),\; x=\frac{m_V^2}{s},
\end{equation}
 and the integral~(\ref{eq:genIv2ImTD}) can be rewritten as
\begin{equation}
\label{eq:genIv2fg}{\cal{I}}_V\sim -2\pi^2\int\limits_{\kappa_0^2}^{s/2} \frac{d\vec{\kappa}^2}{\vec{\kappa}^2}f_g(x,\vec{\kappa}^2,\mu^2),
\end{equation}
which can be finally expressed in terms of the integrated parton distribution $f_g(x,\mu^2)$.

\subsubsection{Extraction of the model parameter in the 3-Pomeron model}
\label{sec:evmp3P1}
As a quantitative example, in this section we consider the case of the modified Regge-eikonal model with three Pomerons~\cite{3P:v1}. Model 
parameters with errors and fits to the data on $J/\Psi$ photoproduction can be found in~\cite{3P:EVMP2005}. It 
is clear from~\cite{3P:EVMP2005} that the main
contribution to the amplitude comes from the ``Born term'' of the so called "hard" ("3rd") Pomeron in the model. Let us write corresponding
amplitudes and use~(\ref{eq:masterEVMP}) to extract the coupling for the gluon-proton amplitude $T^D$:
\begin{eqnarray}
&& \label{eq:EVMPTD} T^D\simeq\eta_{P_3}c^{(3)}_{gp}\mathrm{e}^{B^{(3)}_0 t}\left( \frac{2p\kappa}{s_0-\kappa^2}\right)^{\alpha_{P_3}(t)},\\
&& {\cal{I}}_V=I_V c^{(3)}_{gp} \eta_{P_3}\mathrm{e}^{B^{(3)}_0 t}\left( \frac{s}{s_0}\right)^{\alpha_{P_3}(t)},\\
&& {\cal M}_{i,i}^{V^*+p\to V+p}=c^{(3)}_{Vp} \eta_{P_3}\mathrm{e}^{B^{(3)}_0 t}\left( \frac{s}{s_0}\right)^{\alpha_{P_3}(t)}\\
&& B^{(3)}_0=\frac{1}{4}(\frac{{r_{pP_3}}^2}{2}+{r_{gP_3}}^2),\\
&& \alpha_{P_3}(t)=1+\Delta+\alpha't,\;\Delta=0.2032\pm 0.0041,\; \nonumber\\
&& \alpha'=0.0937\pm 0.0029\;\mathrm{GeV}^{-2},\\
&& {r_{pP_3}}^2=(2.4771\pm 0.0964)\;\mathrm{GeV}^{-2},\; \nonumber\\
&& {r_{gP_3}}^2=(2.54\pm 0.41)\;\mathrm{GeV}^{-2},\\
&& c^{(3)}_{J/\Psi p}=1.11\pm 0.07\;,\; \chi^2/dof=1.48.
\end{eqnarray}
\begin{figure}[bt!]  
  \includegraphics[width=0.49\textwidth]{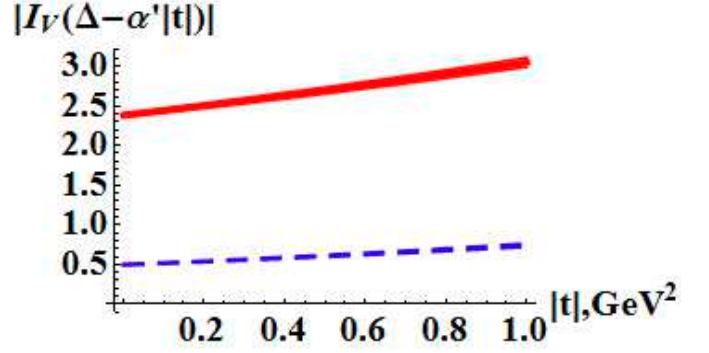} 
  \caption{\label{fig:Iv3PDelta} Function $|I_V(\Delta-\alpha'|t|)|$ versus $|t|$ for $m_V=m_{J/\Psi}=3.1$~GeV (solid) and $m_V=m_{\Upsilon}=9.46$~GeV (dashed).}
\end{figure}
From~(\ref{eq:masterEVMP}) we have
\begin{equation}
\label{eq:cgpextract}c^{(3)}_{gp}=\frac{3\pi^2}{8\alpha_s(m_V^2)\left| I_V\right|}c^{(3)}_{Vp}.
\end{equation}
With rather good accuracy of about 1\% it is possible to calculate $|I_V|$:
\begin{equation}
\left| I_V\right|\simeq \frac{\pi^2}{2^{2+\Delta}}\int\limits_0^1 \frac{dy\; y^{\Delta}(4+y)}{(1+y)(1+y/y_0)^{\Delta+1}},
\end{equation}
where $\Delta=\alpha_{P_3}(t)-1$. $|I_V|$ as a function of $|t|$ is
depicted in the Fig.~\ref{fig:Iv3PDelta}. 
% More exact calculations can be found in the Appendix~A. 

 Since in~\cite{3P:EVMP2005} the fitting procedure was done for the data in the interval $0<|t|<1\;\mathrm{GeV}^2$ we have to
use more general equation~(\ref{eq:masterEVMPt}), which leads to the replacement
\begin{equation}
\label{eq:cgpextracttot}\left| I_V\right|\Longrightarrow <I_V>=\sqrt{\frac{\int\limits_0^1 d|t| \mathrm{e}^{-2B|t|}\left| I_V\right|^2}{\int\limits_0^1 d|t| \mathrm{e}^{-2B|t|}}}
\end{equation}
in~(\ref{eq:cgpextract}), where $B=B^{(3)}_0+\alpha'\ln (s/s_0)$. Finally we get the coupling value
\begin{equation}
\label{eq:cgp3Pv1} c^{(3)}_{gp}=6.535\pm 0.418.
\end{equation}
Here we use another definitions and prescriptions for calculations in comparison with~\cite{3P:EVMP2005}, that is why the value of 
$c_{gp}$ is different. It is more correct than in~\cite{3P:EVMP2005} and further
we will use the definitions and formulae of the present paper for convenience. Errors of $c_{gp}$ are estimated from the errors of all the parameters
in~(\ref{eq:cgpextract}) with the replacement~(\ref{eq:cgpextracttot}).

\subsubsection{Predictions for $J/\Psi$ to $\Upsilon$ production ratio.}
\label{sec:evmp3P2}
Coupling constant $c^{(3)}_{gp}$ have to be the same for any produced vector meson. We can use this fact to check
the model predictions for $J/\Psi$ and $\Upsilon$ mesons. For the total cross-sections ratio it is possible to write
\begin{eqnarray}
{\cal{R}}_{th.}&=&\frac{\sigma_{\gamma p\to\Upsilon p}(W_{\Upsilon})}{\sigma_{\gamma p\to J/\Psi p}(W_{J/\Psi})}\nonumber\\
&\simeq&\left(
\frac{\alpha_s(m_{\Upsilon})^2<I_{\Upsilon}>(m_{\Upsilon},W_{\Upsilon}) W_{\Upsilon}^{\Delta}}{\alpha_s(m_{J/\Psi}^2)<I_{J/\Psi}>(m_{J/\Psi},W_{J/\Psi}) W_{J/\Psi}^{\Delta}}
\right)^2
\nonumber\\
&\times&
\frac{\frac{\Gamma_{\Upsilon\to e^+ e^-}}{m_{\Upsilon}}}{\frac{\Gamma_{J/\Psi\to e^+ e^-}}{m_{J/\Psi}}},\label{eq:Rtheor}
\end{eqnarray}
where
\begin{eqnarray}
&& m_{J/\Psi}=3.1\;\mathrm{GeV},\; \alpha_s(m_{J/\Psi}^2)=0.25, \nonumber \\
&&\label{eq:JPsi} \Gamma_{J/\Psi\to e^+ e^-}=5.52\pm 0.18\;\mathrm{keV}, \\
&& m_{\Upsilon}=9.46\;\mathrm{GeV},\;\alpha_s(m_{\Upsilon}^2)=0.182, \nonumber\\
&& \label{eq:Upsilon}\Gamma_{\Upsilon\to e^+ e^-}=1.34\pm 0.05\;\mathrm{keV}.
\end{eqnarray}

Results of comparison based on the ex\-pe\-ri\-mental data \cite{HERAevmp1}-\cite{HERAevmp2a} are presented in the Table~\ref{tab:RexpHERA}. As you
can see the model adequately describes the data.
%   sigma_J/Psi/sigma_Upsilon ratio (theor. and exp.)
\begin{table}[t!]
\caption{\label{tab:RexpHERA} Theoretical predictions and ex\-pe\-ri\-mental results for the ratio ${\cal{R}}$ at different values of collision energies of $J/\Psi$ and $\Upsilon$ photoproduction. Experimental 
data are taken from~\cite{HERAevmp1}-\cite{HERAevmp2a}.}
\centering
\begin{tabular}{|c|c|c|c|}
\hline
 $W_{J/\Psi}$, GeV & $W_{\Upsilon}$, GeV & ${\cal{R}}_{exp.}\times 10^3$ & ${\cal{R}}_{th.}\times 10^3$\\
 \hline
 $20$-$30$ & $60$-$130$ & $4.91\pm 2.23$ & $3.49\pm 0.64$\\
 \hline
 $20$-$30$ & $130$-$220$ & $9.85\pm 4.37$ & $4.43\pm 0.66$\\
 \hline
 $20$-$30$ & $60$-$220$ & $7.21\pm 2.45$ & $4.06\pm 1.03$\\
 \hline
 $30$-$50$ & $60$-$130$ & $3.86\pm 1.55$ & $2.89\pm 0.56$\\
 \hline
 $30$-$50$ & $130$-$220$ & $7.73\pm 3.0$ & $3.68\pm 0.59$\\
 \hline
 $30$-$50$ & $60$-$220$ & $5.66\pm 1.49$ & $3.37\pm 0.88$\\
 \hline 
 $50$-$70$ & $60$-$130$ & $2.87\pm 1.15$ & $2.47\pm 0.44$\\
 \hline
 $50$-$70$ & $130$-$220$ & $5.75\pm 2.24$ & $3.13\pm 0.44$\\
 \hline
 $50$-$70$ & $60$-$220$ & $4.21\pm 1.12$ & $2.87\pm 0.72$\\
 \hline 
 $70$-$90$ & $60$-$130$ & $2.4\pm 0.99$ & $2.2\pm 0.38$\\
 \hline
 $70$-$90$ & $130$-$220$ & $4.82\pm 1.9$ & $2.79\pm 0.37$\\
 \hline
 $70$-$90$ & $60$-$220$ & $3.53\pm 0.96$ & $2.56\pm 0.63$\\
 \hline 
 $90$-$110$ & $60$-$130$ & $2.18\pm 0.88$ & $2.01\pm 0.34$\\
 \hline
 $90$-$110$ & $130$-$220$ & $4.37\pm 1.7$ & $2.56\pm 0.33$\\
 \hline
 $90$-$110$ & $60$-$220$ & $3.2\pm 0.85$ & $2.34\pm 0.57$\\
 \hline 
 $110$-$130$ & $60$-$130$ & $1.85\pm 0.74$ & $1.87\pm 0.31$\\
 \hline
 $110$-$130$ & $130$-$220$ & $3.7\pm 1.44$ & $2.38\pm 0.3$\\
 \hline
 $110$-$130$ & $60$-$220$ & $2.71\pm 0.71$ & $2.18\pm 0.53$\\
 \hline 
 $130$-$150$ & $60$-$130$ & $1.54\pm 0.63$ & $1.76\pm 0.29$\\
 \hline
 $130$-$150$ & $130$-$220$ & $3.09\pm 1.23$ & $2.24\pm 0.28$\\
 \hline
 $130$-$150$ & $60$-$220$ & $2.26\pm 0.63$ & $2.05\pm 0.5$\\
 \hline 
 $150$-$170$ & $60$-$130$ & $1.46\pm 0.61$ & $1.67\pm 0.28$\\
 \hline
 $150$-$170$ & $130$-$220$ & $2.92\pm 1.19$ & $2.12\pm 0.27$\\
 \hline
 $150$-$170$ & $60$-$220$ & $2.14\pm 0.62$ & $1.94\pm 0.47$\\
 \hline  
\end{tabular}
\end{table}

\section{Exclusive double diffractive events.}

 In this section we consider the exclusive central diffraction. The previous simple
 model with three Pomerons was presented in~\cite{newmodels5}-\cite{ouroldhiggs}. Here
 is the more general approach.

\subsection{Kinematics.}
\label{sec:eddeKIN}

\begin{figure}[t!]  
  \includegraphics[width=0.49\textwidth]{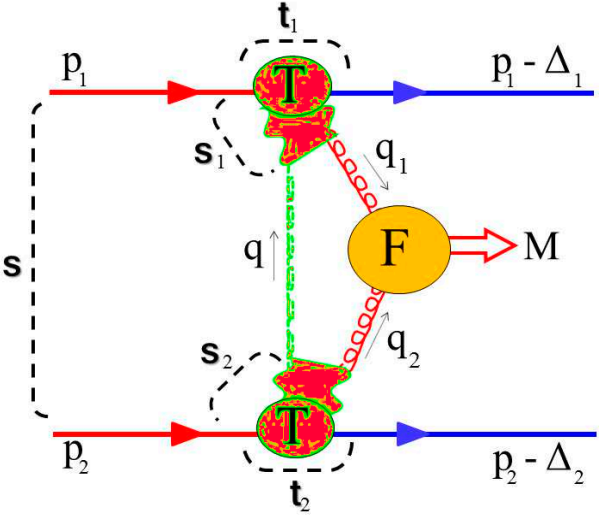} 
  \caption{\label{fig:EDDEkin} ``Bare'' amplitude and kinematics of the Exclusive Double Diffractive Event (EDDE) in the case of perturbative Pomeron-Pomeron 
  fusion. Absorbtion in the initial and final pp channels is not shown.}
\end{figure}

Let us remind to the reader some kinematical formulae. In the 
Fig.~\ref{fig:EDDEkin} we illustrate the perturbative mechanism for the ``bare'' amplitude of the process
$p+p\to p+M+p$. We use the kinematics, which corresponds 
to the double Regge limit. It is convenient
to use light-cone representation for momenta. The notations are
\begin{eqnarray}
s&=&(p_1+p_2)^2,\; s^{\prime}=(p_1^{\prime}+p_2^{\prime})^2,\nonumber\\
\bar{s}&=&p_1p_2+\sqrt{(p_1p_2)^2-m^4}=\frac{s-2m^2}{2}+\frac{s}{2}\sqrt{1-\frac{4m^2}{s}},\nonumber\\
p_1&=&\left\{ \sqrt{\frac{\bar{s}}{2}},\frac{m^2}{\sqrt{2\bar{s}}};\; \vec{0}\right\},\;
p_2=\left\{ \frac{m^2}{\sqrt{2\bar{s}}},\sqrt{\frac{\bar{s}}{2}};\; \vec{0}\right\},\nonumber\\
p_1^{\prime}&=&p_1-\Delta_1=
\left\{ 
(1-\xi_1)\sqrt{\frac{\bar{s}}{2}},\frac{\vec{\Delta}_1^2+m^2}{(1-\xi_1)\sqrt{2\bar{s}}};\; -\vec{\Delta}_1
\right\},\nonumber\\
p_2^{\prime}&=&p_2-\Delta_2=\left\{
\frac{\vec{\Delta}_2^2+m^2}{(1-\xi_2)\sqrt{2\bar{s}}},(1-\xi_2)\sqrt{\frac{\bar{s}}{2}};\; -\vec{\Delta}_2
\right\},\nonumber\\
\tilde{p}_1&=& p_1-\frac{m^2}{\bar{s}}p_2= \left\{ \sqrt{\frac{\bar{s}}{2}}
\left(1-\frac{m^4}{\bar{s}^2}\right),0;\; \vec{0}\right\},\;\nonumber\\
\tilde{p}_2&=& p_2-\frac{m^2}{\bar{s}}p_1= \left\{ 0, \sqrt{\frac{\bar{s}}{2}}
\left(1-\frac{m^4}{\bar{s}^2}\right);\; \vec{0}\right\},\nonumber\\
p_{1,2}^2&=&p_{1,2}^{\prime\; 2}=m^2,\; \tilde{p}_{1,2}^{\; 2}=0,
\label{eq: EDDEmomenta0}
\end{eqnarray}
\begin{eqnarray}
q&= & \sqrt{\frac{2}{s}}(q_+\tilde{p}_1+q_-\tilde{p}_2)+q_{\perp}=\left\{ q_+, q_-;\; \vec{q}\right\}\;{,}\nonumber\\
\label{eq: EDDEmomenta}
q_1&= &q+\Delta_1,\; q_2=-q+\Delta_2,\; 
\Delta_{1,2}\simeq\xi_{1,2}\tilde{p}_{1,2},\; 
\end{eqnarray}
where $\xi_{1,2}$ are fractions of protons' momenta carried by gluons. From the above notations we can
obtain the relations:
\begin{eqnarray}
t_{1,2}&=&\Delta_{1,2}^2 \simeq -\frac{\vec{\Delta}_{1,2}^2+\xi_{1,2}^2m^2}{1-\xi_{1,2}}\;\;\simeq
-\vec{\Delta}_{1,2}^2,\;\xi_{1,2}\to 0\nonumber\\
M^2&=& (q_1+q_2)^2\simeq \xi_1\xi_2s+t_1+t_2-2\sqrt{t_1t_2}\cos\phi_0\nonumber\\
M_{\perp}^2&=& \xi_1\xi_2s \simeq M^2+|t_1|+|t_2|+2\sqrt{t_1t_2}\cos\phi_0\nonumber\\
&&\cos\phi_0=\frac{\vec{\Delta}_1\vec{\Delta}_2}{|\vec{\Delta}_1||\vec{\Delta}_2|}\label{eq:EDDEnotations1a}\\
s_1&=&(p_1+q)^2\simeq m^2+q^2+\sqrt{2s}q_-\nonumber\\
\label{eq:EDDEnotations1b}s_2&=&(p_2-q)^2\simeq m^2+q^2-\sqrt{2s}q_+\; .
\end{eqnarray}
Physical region of diffractive events with two rapidity gaps is defined by the following  
kinematical cuts:
\begin{eqnarray}
\label{eq:tlimits}
&&0.01\; GeV^2\le |t_{1,2}|\le\; \sim 1\; GeV^2\;{,} \\
&&\label{eq:xilimits}
\xi_{min}\simeq\frac{M^2}{s \xi_{max}}\le \xi_{1,2}\le \xi_{max}\sim 0.1\;,\\
\label{eq:kappalimits}
&&\left(\sqrt{-t_1}-\sqrt{-t_2}\right)^2\le\kappa\le\left(\sqrt{-t_1}+\sqrt{-t_2}\right)^2\\
&&\kappa=\xi_1\xi_2s-M^2\ll M^2\nonumber
\end{eqnarray}
We can write the relations in terms of $y_{1,2}$ and $y$ (rapidities of hadrons and the system M correspondingly). For instance:
\begin{eqnarray}
&&\xi_{1,2}\simeq\frac{M}{\sqrt{s}}e^{\pm y},\;
|y|\le y_0=\ln\left(\frac{\sqrt{s}\xi_{max}}{M}\right),\nonumber\\
\label{eq:raplimits}&&|y_{1,2}|=\frac{1}{2}\ln\frac{(1-\xi_{1,2})^2s}{m^2-t_{1,2}}\ge 9.
\end{eqnarray}
%$y_0\simeq 2.5$ for $\sqrt{s}=14\; TeV$ and $M=O(100\; GeV)$.

\subsection{Basic ingredients.}

\begin{figure}[h!]  
 \includegraphics[width=0.49\textwidth]{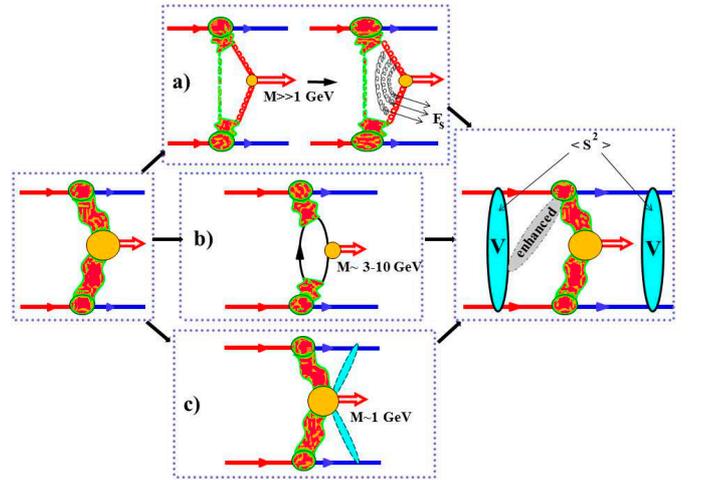} 
  \caption{\label{fig:EDDEdyn} Scheme of calculation of the full EDDE amplitude
  in the case of high (a), intermediate (b) and low (c) invariant mass, i.e. perturbative, intermediate 
  and nonperturbative
  Pomeron-Pomeron fusion correspondingly.}
\end{figure}

Let us consider basic constituents of the model framework. We begin
from the case of perturbative representation of the Pomeron-Pomeron fusion as
depicted in Fig.~\ref{fig:EDDEkin} and in
Fig.~\ref{fig:EDDEdyn}a. In the perturbative case we have to use proton-gluon amplitudes $T$ convoluted with gluon-gluon fusion amplitude $F$ (see Fig.~\ref{fig:EDDEkin}), then 
take into account Sudakov-like suppression (depicted by curved wavy lines in Fig.~\ref{fig:EDDEdyn}a), and, finally, absorbtive (or rescattering) corrections denoted by $V$-blobs. In 
the nonperturbative case it is not possible to consider
Pomeron as a singlet two-gluon state, and we also have to take into account
the interaction between the central low mass state with the protons (blobs with the dashed boundary in the Fig.~\ref{fig:EDDEdyn}c). Recently it was
shown in~\cite{newmodels4} that enhanced diagrams (additional soft interactions)
can play significant role.

\subsubsection{Proton-gluon scattering.}

Diffractive part $T$ of the amplitude is calculated the same way
as in~(\ref{eq:T}):
\begin{equation}
\label{eq:T12}T^{\alpha\mu}_1\equiv 
 T^{\alpha\mu}(p_1,-q,q_1),\; 
 T^{\alpha\nu}_2\equiv 
 T^{\alpha\nu}(p_2,q,q_2)
\end{equation}
Here we take apart color delta-functions in all amplitudes, which give overall 
factor $8$ after contractions and neglect terms of 
the order $o(\xi_i)$, $o(t_i/m^2)$. Equations~(\ref{eq:Tgaugeinvar}) are used
to simplify calculations.

\subsubsection{Gluon-gluon fusion.}

The main contribution to the gluon-gluon fusion tensor can be represented as
\begin{equation}
\label{eq:Fgg}
F_{\mu\nu}=G_{\mu\nu} F_{gg\to M},\; 
G_{\mu\nu}= g_{\mu\nu}-\frac{q_{2,\;\mu}q_{1,\;\nu}}{q_1q_2},
\end{equation}
where $F_{gg\to M}$ is related to the differential cross-section
\begin{eqnarray}
\frac{d\hat{\sigma}_{gg\to M}}{d\hat{t}}&=&
\frac{1}{(2s_g^{(1)}+1)(2s_g^{(2)}+1)N_g^2}N_g\nonumber\\
&\times&\left| F_{gg\to M}\right|^2
\frac{\sum\limits_{i=1}^2 \left| G_{\mu\nu}\epsilon_i^{\mu}\epsilon_i^{\nu}\right|^2}{16\pi M^4},
\end{eqnarray}
where
\begin{eqnarray}
&&N_g=N_c^2-1=8,\; 2s_g^{(i)}+1=2\; (q_i^2=0),\;\nonumber\\
&&\left| F_{gg\to M}\right|^2=256\pi M^4\frac{d\hat{\sigma}_{gg\to M}}{d\hat{t}},\nonumber
\end{eqnarray}
and $M=gg,\; Q\bar{Q},\; \gamma\gamma,\; W^+W^-,\; ZZ$ or other 
two-particle system with final momenta $k_{1,2}$ and $\hat{t}=(q_1-k_1)^2$. For
one particle production ($M=H,\; G\mbox{(graviton)},\; \chi_{b,c}$) $F_{gg\to M}$ can
be expressed in terms of the hadronic decay width of a particle
\begin{equation}
\left| F_{gg\to M}\right|^2=64\pi^2\frac{\Gamma_{M\to gg}}{M}.
\end{equation}
Here we neglect other transverse tensor structure in $F_{\mu\nu}$ 
proportional to $q_{1,\;\mu}q_{2,\;\nu}$ since
it gives small contribution to the result.

\subsubsection{Sudakov-like suppression.}

In the perturbative case there is an additional factor for the gluon-gluon fusion vertex, which
is depicted as virtual gluon corrections in Fig.~\ref{fig:EDDEdyn}a. If we take 
into account the emission of virtual "soft" gluons, while prohibiting the real ones, that 
could fill rapidity gaps, this results in
the Sudakov-like suppression~\cite{KMR:sudakov}:

\begin{eqnarray}
&&F_s(l^2,\mu^2)= \exp
\left[ -
\int\limits_{l^2}^{{\mu}^2} \frac{d p_T^2}{p_T^2} \,
\frac{\alpha_s({p_T}^2)}{2\pi} \int\limits_{\Delta_{(\mu)}}^{1-\Delta_{(\mu)}} z P_{gg}(z) d z
\right.
\nonumber\\
&&\hspace*{2.8cm}+\left.\phantom{\int\limits_{l^2}^{{\mu}^2}}\hspace*{-0.5cm}\int\limits_0^1\sum_q P_{qg}(z)dz
\right].\label{eq:sudakov}
\end{eqnarray}
Here
\begin{eqnarray}
&& P_{gg}(z) = 6\frac{(1-z(1-z))^2}{z(1-z)},\;\nonumber\\
&& \Delta_{(\mu)} =\frac{p_T}{p_T+\mu},\; l^2=-q^2,\; \mu=M/2,
\end{eqnarray}
and $q$ is the loop momentum in Fig.~\ref{fig:EDDEkin}.

\subsubsection{Soft rescattering.}

We have to take into account also the hadron-hadron ``soft'' interaction in the initial and in the final 
states (unitary corrections or re-scattering). It is denoted by $V$ in Fig.~\ref{fig:EDDEdyn} and given 
by the following analytical
expressions:
\begin{eqnarray}
&&{\cal M}^{U}(p_1, p_2, \Delta_1, \Delta_2) = \int \frac{d^2\vec{q}_T}{(2\pi)^2} \,
\frac{d^2\vec{q}^{\;\prime}_T}{(2\pi)^2} \; V(s, \vec{q}_T) \;
\nonumber \\
\label{ucorr}&&\times {\cal M}( p_1-q_T, p_2+q_T,\Delta_{1T}, \Delta_{2T}) \,
V(s^{\prime}, \vec{q}^{\;\prime}_T) \;,
\\
&&V(s, \vec{q}_T) = \int d^2\vec{b} \, e^{i\vec{q}_T
\vec{b}} \sqrt{1+2\mathrm{i} T^{el}_{pp\to pp}(s, \vec{b})},
\end{eqnarray}
where $\Delta_{1T} = \Delta_{1} -q_T - q^{\prime}_T$, $\Delta_{2T}
= \Delta_{2} + q_T + q^{\prime}_T$, ${\cal M}$ is the ``bare'' amplitude of the
process $p+p\to p+M+p$. In the case of the eikonal representation of the
elastic amplitude $T^{el}_{pp\to pp}$ we have
\begin{equation}
\label{eq:Veik}
V(s, \vec{q}_T) = \int d^2\vec{b} \, \mathrm{e}^{\mathrm{i}\vec{q}_T
\vec{b}} \mathrm{e}^{\mathrm{i}\delta_{pp\to pp}(s,\vec{b})},
\end{equation}
where $\delta_{pp\to pp}$ is the eikonal function. As was shown in
\cite{spin_parity_analyser4}, these ``outer'' unitary corrections
strongly reduce the value of the corresponding cross-section and
change the azimuthal angle dependence. The ratio
\begin{equation}
\label{eq:SoftSurv0}
<S^2>=\frac{\int\int d^2\vec{\Delta}_1 d^2\vec{\Delta}_2 \left| {\cal M}^{U}\right|^2}{\int\int d^2\vec{\Delta}_1 d^2\vec{\Delta}_2 \left| {\cal M}\right|^2}
\end{equation}
is usually called ``soft survival probability''.

\subsubsection{Full construction. Convolution and integration. EDDE luminocity.}

Let us collect all the ingredients of the framework. First of all we
calculate the loop integral
\begin{eqnarray}
{\cal I}_q &=& 2\pi\int\frac{d^4q}{(2\pi)^4}
\frac{f(q_+,q_-,\vec{q}^{\;2},\dots)}{\left( q^2+\mathrm{i}0\right)\left( q_1^2+\mathrm{i}0\right)\left( q_2^2+\mathrm{i}0\right)}\;,\nonumber\\
q_{1\atop 2}^2+\mathrm{i}0 &\simeq & \pm\sqrt{2s}\;\xi_{1 \atop 2}\left( q_{\mp} \pm \left( \frac{\vec{q}_{1\atop 2}^{\;2}}{\sqrt{2s}\;\xi_{1\atop 2}} -\mathrm{i}0\right)\right)\label{eq:Iq1},
\end{eqnarray}
factor $2\pi$ before the integral is introduced for convenience.
It was shown in~\cite{Collins}, that the leading contribution arises from the region of the
integration, where momentum $q$ is "Glauber-like", i.e. of the order 
$(\mbox{k}_+m^2/\sqrt{s},\mbox{k}_-m^2/\sqrt{s},\mbox{\bf k}m)$, where k's are of the 
order $1$. The detailed consideration of the loop integral 
shows that the main contribution comes from the poles at $q_i^2=0$
\begin{eqnarray}
{\cal I}_q &\!\!=&\!\! \frac{1}{2^4 M^2} \int\limits_0^{\frac{M^2}{4}}\frac{d\vec{q}^{\;2} f
(
-\frac{\vec{q}_2^{\;2}}{\sqrt{2s}\xi_2}, \frac{\vec{q}_1^{\;2}}{\sqrt{2s}\xi_1}, \vec{q}^{\;2},\dots 
)}{
\left( \vec{q}^{\;2} + \frac{\vec{q}_1^{\;2}\vec{q}_2^{\;2}}{M^2}\right)}\;,
\nonumber\\
\label{eq:Iq2}\vec{q}_{1\atop 2}^{\;2} &\!\!=&\!\! \vec{q}^{\;2}+\vec{\Delta}_{1\atop 2}^{\;2}\pm2|\vec{q}||\vec{\Delta}_{1\atop 2}|\cos(\phi\pm\frac{\phi_0}{2})\;.
\end{eqnarray}
As to the lower limit in the integral~(\ref{eq:Iq2}), we set it to zero since
$f|_{\vec{q}^{\;2}=0}=0$ and the main contribution comes from the region $\vec{q}_i^{\;2}/\xi_i>m^2$, which gives $\vec{q}^2\; >\; <\!\!|t_i|\!\!>\sim 0$.

\noindent In our case we have to replace the function $f$ in~(\ref{eq:Iq2}) by
\begin{eqnarray}
\tilde{f}&=&8 F_{\mu\nu}T_1^{\alpha\mu}T_2^{\alpha\nu}F_s(-q^2,\frac{M^2}{4})\nonumber\\
&\simeq& 
8\frac{2M^2 \vec{q}_1\vec{q}_2}{\vec{q}_1^{\;2}\vec{q}_2^{\;2}} F_s(\vec{q}^{\;2},\frac{M^2}{4}) T^D_1 T^D_2F_{gg\to M},\nonumber\\
T^D_{1\atop 2}&\equiv& T^D\left(\frac{\sqrt{s}}{M}\mathrm{e}^{\mp y}\vec{q}_{1\atop 2}^{\;2}\sqrt{1-\frac{M}{\sqrt{s}}\mathrm{e}^{\pm y}}\right)
\label{eq:FTDTD}
\end{eqnarray}

 Finally to obtain the EDDE cross-section we can introduce the luminocity function
\begin{equation}
\hat{\cal L}_{EDDE}=\frac{1}{2^9\pi^6}\left( \frac{M^2}{s}\right)^2 \left| {\cal I}_q\right|^2 <S^2>,
\label{eq:LumEDDE}
\end{equation}
where for low values of $\vec{\Delta}_{1,2}$ we replace $\vec{q}_{1,2}$ by $\vec{q}$ in ${\cal I}_q$. Now we can write
\begin{equation}
\label{eq:EDDEcs}
M^2\frac{d\sigma_{EDDE}}{dM^2\; dy \;d\Phi_{gg\to M}}=\hat{\cal L}_{EDDE} \frac{d\hat{\sigma}_{gg\to M}^{J_z=0}}{d\Phi_{gg\to M}}.
\end{equation}
For resonance production Eq.~(\ref{eq:EDDEcs}) can be simplified to
\begin{equation}
\label{eq:EDDEcsRes}
\frac{d\sigma^{Res}_{EDDE}}{dy}=\hat{\cal L}_{EDDE} \frac{2\pi^2\Gamma_{M\to gg}}{M^3},
\end{equation}
where $\Gamma_{M\to gg}$ is the resonance width.

As in the case of EVMP we can use any model for $T^D$. For example, 
\begin{equation}
T^D(\frac{\;\vec{q}^{\;2}}{\xi}\sqrt{1-\xi})\Longrightarrow f_g(\xi,\vec{q}^{\;2},\mu^2).
\label{eq:EDDEpartonicTD}
\end{equation}

Gluon-gluon fusion functions can be found in the Appendix A.

\subsubsection{3-Pomeron model as an example.}

\begin{figure}[h!]  
  \includegraphics[width=0.49\textwidth]{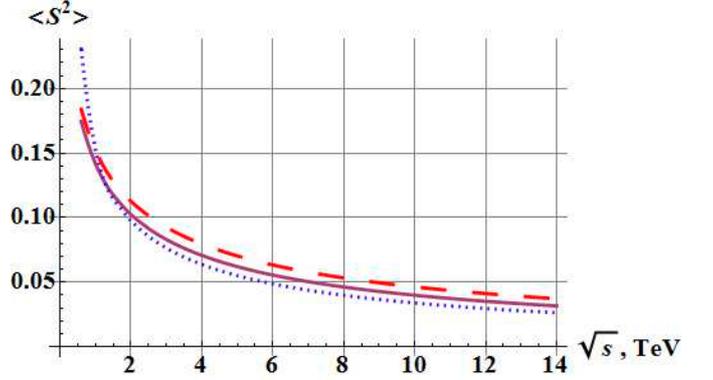} 
  \caption{\label{fig:S2} ``Soft survival probability'' in the case of the 3-Pomeron model for different values
  of the invariant mass: $M=30$~GeV (dashed), $M=125$~GeV (solid) and $M=600$~GeV (dotted).}
\end{figure}

\begin{figure}[hbt] 
  \includegraphics[width=0.49\textwidth]{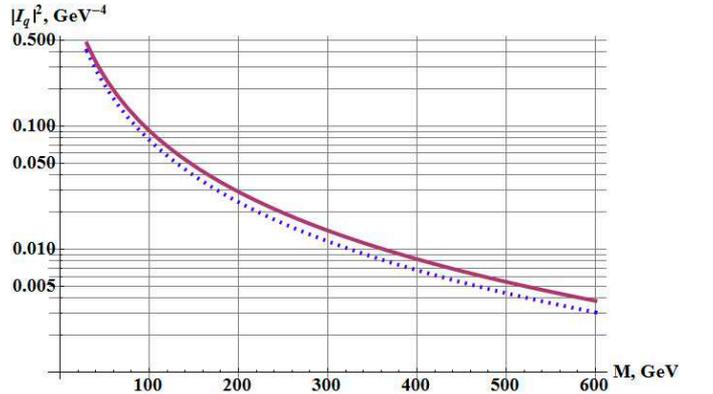} 
  \caption{\label{fig:Iq} Function $|I_q|^2$ versus $M$ for different values of the parameter $\Delta$ in the case of the 3-Pomeron 
  model. $\Delta=\alpha_{P_3}(0)-1$ (solid) and $\Delta=\alpha_{P_3}(1)-1$ (dotted).}
\end{figure}

\begin{figure}[hbt]  
 \includegraphics[width=0.49\textwidth]{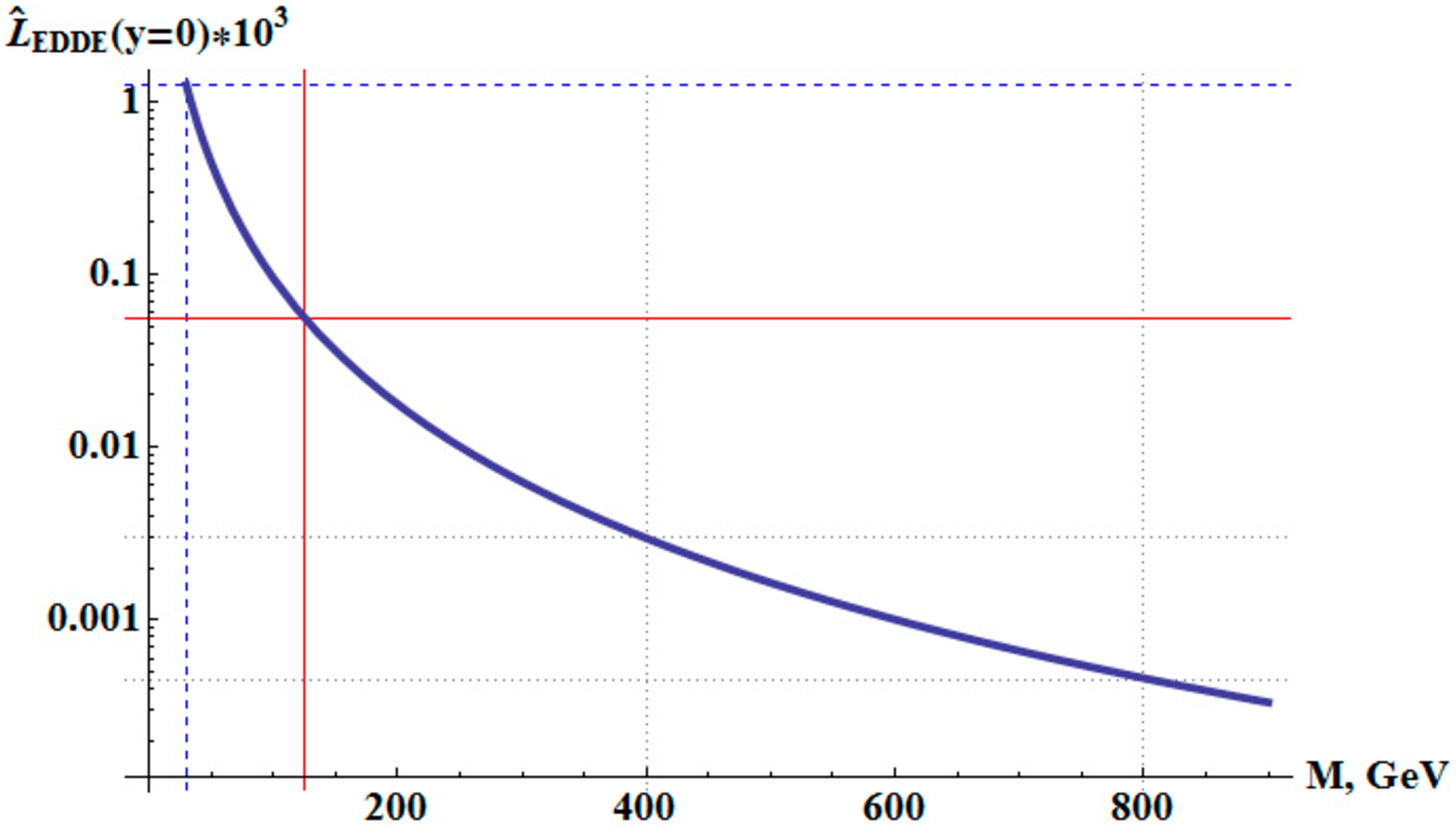} \includegraphics[width=0.49\textwidth]{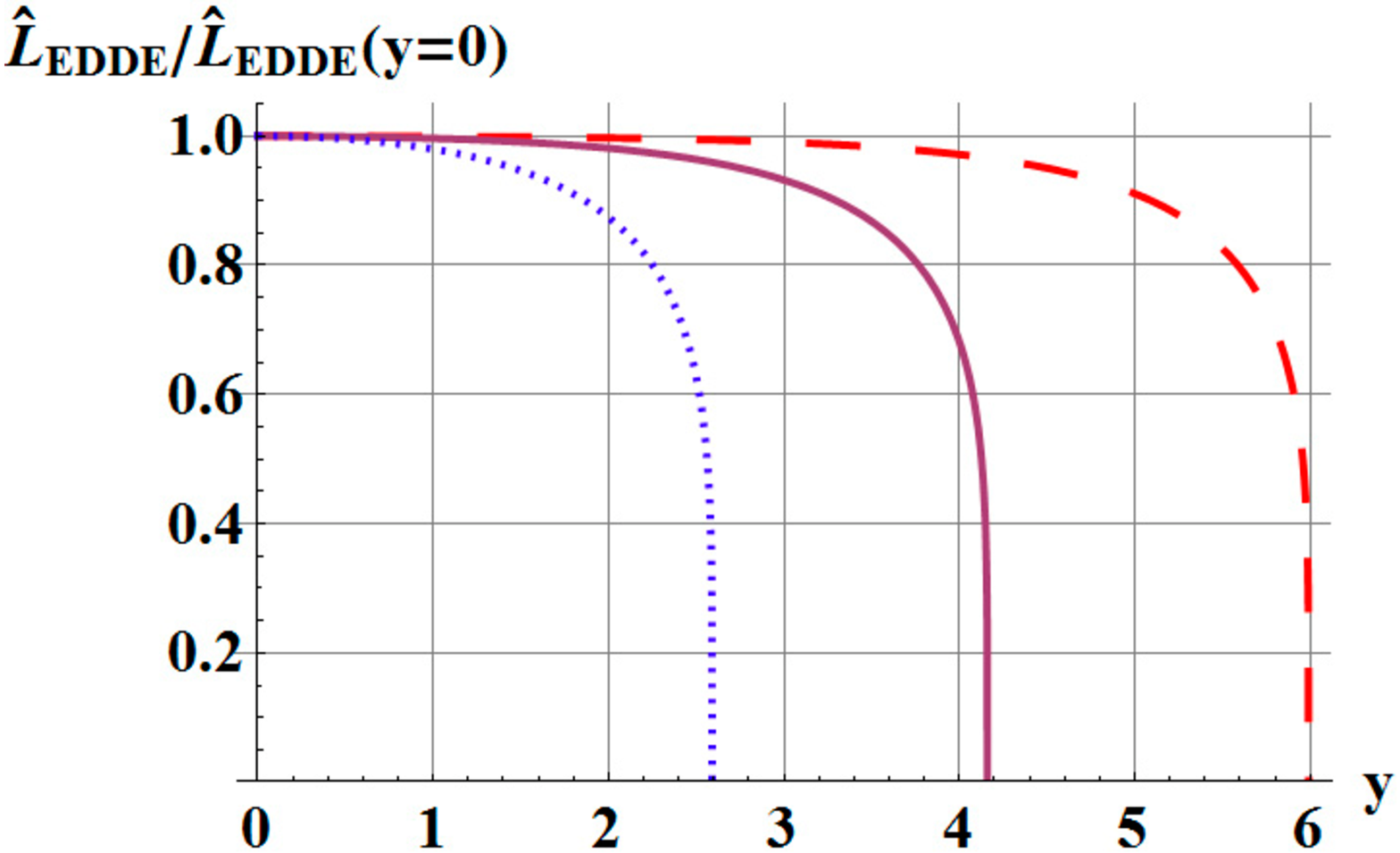}
  \caption{\label{fig:LEDDE} Pomeron-Pomeron luminocity $\hat{\cal L}_{EDDE}$ at $y=0$ (top picture) 
and as a function of $y$ for different fixed invariant masses (bottom picture)
in the case of the 3-Pomeron model. In the left picture solid grid line denotes $M=125$~GeV and dashed one denotes 
$M=30$~GeV. In the right picture curves correspond to 
$M=20$~GeV (dashed), $M=125$~GeV (solid) and $M=600$~GeV (dotted).}
\end{figure}

In the 3-Pomeron model the diffractive proton-gluon amplitude is similar to~(\ref{eq:EVMPTD})
\begin{equation}
\label{eq:EDDE3PTD}
T^D_i\simeq\eta_{P_3}c^{(3)}_{gp}\mathrm{e}^{B^{(3)}_0 t_i}\left( \frac{s_i-m^2-qq_i}{s_0-qq_i}\right)^{\alpha_{P_3}(t_i)},
\end{equation}
and for the EDDE luminocity we have
\begin{eqnarray}
&&\hat{\cal L}_{EDDE}=\frac{\left| \eta_{P_3}c^{(3)}_{gp}\right|^4}{2^9\pi^6}
\frac{1}{4B^2} \left( \frac{s}{M^2}\right)^{2\Delta} \nonumber\\
&&\times\left( 1-\frac{2M}{\sqrt{s}}\cosh y +\frac{M^2}{s}\right)^{\Delta}
\left| {I}_q\right|^2 <S^2>,\label{eq:LumEDDE3P}
\end{eqnarray}
where
\begin{eqnarray}
&&I_q=\int\limits_{<|t_i|>}^{\frac{M^2}{4}} \frac{d\vec{q}^{\;2}}{\vec{q}^{\;2}}(\vec{q}_1\vec{q}_2) 
\nonumber\\
&&\times \frac{(\vec{q}_1^{\;2})^{\alpha_{P_3}(t_1)-1}
(\vec{q}_2^{\;2})^{\alpha_{P_3}(t_2)-1}}{
\left( s_0+\vec{q}^{\;2}/2\right)^{\alpha_{P_3}(t_1)+\alpha_{P_3}(t_2)}} F_s(\vec{q}^{\;2},\frac{M^2}{4}) \nonumber\\
&& \label{eq:EDDEIq3P}\simeq -\int\limits_{0}^{\frac{M^2}{4}}d\vec{q}^{\;2}
\frac{(\vec{q}^{\;2})^{2\Delta}}{
\left( s_0+\vec{q}^{\;2}/2\right)^{2(1+\Delta)}}
F_s(\vec{q}^{\;2},\frac{M^2}{4}),
\end{eqnarray}
and the ``soft survival probability'' can be reduced to the simple form (see Appendix B):
\begin{eqnarray}
\label{eq:EDDESurv3P}
&& <S^2>\simeq \frac{1}{4B}\int\limits_0^{\infty} \left| h(\tau)\right|^2 d\tau^2,\\
&& \label{eq:EDDESurv3Ph} h(\tau)=\int\limits_0^{\infty} b\; db\; J_0(b \tau) \mathrm{e}^{-\Omega(s,b)-\Omega(s',b)-b^2/(8B)},
\end{eqnarray}
where $B=B^{(3)}_0+\alpha'\ln (\sqrt{s}/M)$, $\Omega\equiv \mathrm{i}\delta_{pp\to pp}$, and $\delta_{pp\to pp}$ can be found in~\cite{3P:v1}. Functions $<S^2>$, $|I_q|^2$ and $\hat{\cal L}_{EDDE}$ are 
presented in Figs~\ref{fig:S2}, \ref{fig:Iq}, \ref{fig:LEDDE}.

\subsubsection{Nonperturbative mechanism of Pomeron-Pomeron fusion. Discussion.}

If the central mass produced in EDDE is low (about several GeV), it is not
possible to use perturbative representation (see Fig.~\ref{fig:EDDEkin} and Fig.~\ref{fig:EDDEdyn}a) 
for the amplitude of the process, and we have to use more general ``nonperturbative'' form (see Fig.~\ref{fig:EDDEdyn}c). In this case we have to obtain somehow the 
vertex ``Pomeron-Pomeron-Central system''. 

From the first principles we can write the general structure of the vertex for
different cases~\cite{spin_parity_analyser4}. For example, for the production
of the system with $J^P$ (spin-parity) we have
\begin{eqnarray}
&& \label{eq:lowm1}\left|{\cal M}^{0^+}\right|^2\sim  (M_{\perp}^2)^{2(\alpha_{\mathbb P}(0)-1)}(f_0M_{\perp}^2+2f_1)^2, \\
&& \label{eq:lowm2}\left|{\cal M}^{0^-}\right|^2\sim  (M_{\perp}^2)^{2(\alpha_{\mathbb P}(0)-1))}f_0 t_1 t_2 \sin^2\phi_0, \\
&& \left|{\cal M}^{1^+}\right|^2\sim (M_{\perp}^2)^{2(\alpha_{\mathbb P}(0)-1)}({\cal F}_0 M_{\perp}^4+{\cal F}_1 t_1 t_2 \sin^2\phi_0+{\cal F}_2),\; \nonumber\\
&&\label{eq:lowm3}{\cal F}_{0,2}\sim o(t_i),\\
&& \label{eq:lowm4}\left|{\cal M}^{2^+}\right|^2\sim (M_{\perp}^2)^{2(\alpha_{\mathbb P}(0)-1)}({\cal F}_0 M_{\perp}^4+{\cal F}_1 M_{\perp}^2+{\cal F}_2),
\end{eqnarray}
with notations~(\ref{eq:EDDEnotations1a}),(\ref{eq:EDDEnotations1b}) and functions defined in~\cite{spin_parity_analyser4}. The general structure
of helicity amplitudes from the simple Regge behaviour was also considered in~\cite{spin_parity_analyser2a},\cite{spin_parity_analyser2b}. Experimental 
data are in good agreement with these predictions.

There are some attempts to obtain the vertex in special models for the Pomeron. In 
Refs.~\cite{spin_parity_analyser3a},\cite{spin_parity_analyser3b} results
were obtained from the assumption that the Pomeron acts as a $1^+$ 
conserved or nonconserved current. The Pomeron-Pomeron fusion based on the ``instanton'' or ``glueball''
dynamics was considered in~\cite{instanton1}-\cite{instanton4}. You can see also recent papers~\cite{holographicPom},\cite{lowmassmesonsLHC}
devoted to calculations of the Pomeron-Pomeron fusion vertex in the nonperturbative regime.

\subsection{Standard candles for high invariant masses.}

One of the basic tasks now is to make predictions for EDDE with
production of fundamental particles like the Higgs boson. First of all
we can check any theoretical model by the use of the
recent data from Tevatron~\cite{CDFreview}-\cite{CDFhq}. ``Standard
candles'' in the case of high invariant mass are exclusive central di-jet and di-gamma production. We can
use also $\chi_{c,0}$ production, but $m_{\chi_{c,0}}=3.5$~GeV and we have to take into account also nonperturbative mechanisms 
of the Pomeron-Pomeron fusion.

\begin{figure}[b!] 
  \includegraphics[width=0.49\textwidth]{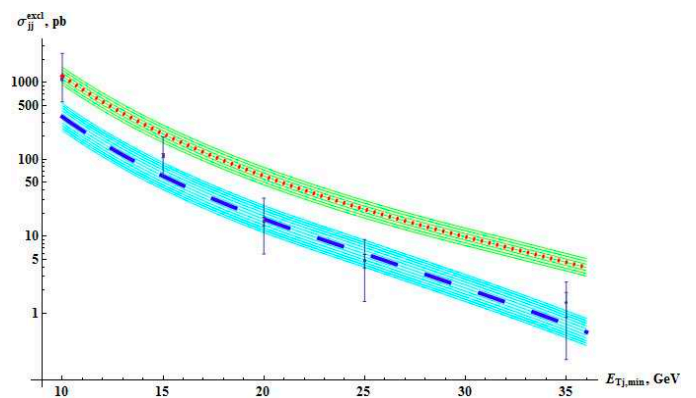} 
  \caption{\label{fig:CDFjj} CDF data on the exclusive di-jet production cross-section~\cite{CDFjj} versus the lower cut of $E_{T,jet}$ and the prediction of the 3P-model. Upper 
  dotted curve corresponds to the value of $c_{gp}^{(3)}$ presented in~(\ref{eq:cgp3Pv1}) and obtained from fitting the HERA data on EVMP. Lower dashed curve is 
  obtained by rescaling $E_{T,jet}=0.75 E_{T,g}$ as was proposed in~\cite{newmodels2},\cite{newmodels3}. Filled areas denote errors from $c_{gp}^{(3)}$.}
\end{figure}

 Let us analyse the CDF data in the framework of the 3-Pomeron model. In Fig.~\ref{fig:CDFjj} you
 can see the prediction based on the HERA data (upper dotted curve) with the value of $c_{gp}^{(3)}$ presented 
 in~(\ref{eq:cgp3Pv1}). As was claimed in~\cite{newmodels2},\cite{newmodels3}, the uncertainty on the hadronic level can be taken into account
 by rescaling of jet $E_T$. After application of this procedure to our result with $E_{T,jet}=0.75 E_{T,g}$ we obtain 
 the lower dashed curve in Fig.~\ref{fig:CDFjj}, which is in quite a good agreement with the CDF data.

\begin{figure}[t!] 
\includegraphics[width=0.49\textwidth]{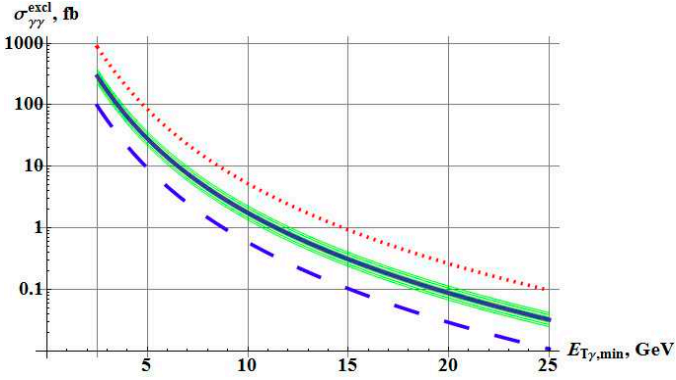} 
  \caption{\label{fig:CDFgamgam} CDF data on the exclusive di-gamma production cross-section~\cite{CDFjj} versus lower cut of $E_{T,\gamma}$ and prediction of the 3P-model. Solid curve
  corresponds to the value of $c_{gp}^{(3)}$ presented in~(\ref{eq:cgp3Pv1}) and obtained from fitting the HERA data on EVMP. Filled areas denote errors from $c_{gp}^{(3)}$. Results
  multiplied by 3 (dotted curve) and devided by 3 (dashed curve) are also shown.}
\end{figure}

The prediction for the exclusive central di-gamma production is shown in Fig.~\ref{fig:CDFgamgam}. The value
of the predicted cross-section after CDF cuts are 
\begin{eqnarray}
\label{3PCDFgamgam5}&&E_T>5\;{\mathrm GeV},\; |\eta_{\gamma}|<1\Longrightarrow  \sigma_{\gamma\gamma}^{excl,\; th}=28\pm 8\;{\mathrm fb},\\
&&E_T>2.5{\mathrm GeV},\; |\eta_{\gamma}|<1\nonumber\\
\label{3PCDFgamgam2}&&\hspace*{2.6cm}\;\Longrightarrow  \sigma_{\gamma\gamma}^{excl,\; th}=0.29\pm 0.08\;{\mathrm pb},
\end{eqnarray}
which are close to the predictions~\cite{newmodels4}. As was shown in~\cite{newmodels4}, uncertainties in the gluon distributions can
lead to a factor ${\times 3}\atop{\div 3}$ (upper and lower curves are shown for the illustration of this fact in the Fig.~\ref{fig:CDFgamgam}) in 
the theoretical prediction. If we take into account this fact, CDF results~\cite{CDF2gamA},\cite{CDF2gamB}
\begin{eqnarray}
\label{CDF2gam5}&& \hspace*{-1cm}E_T>5\;{\mathrm GeV},\; |\eta_{\gamma}|<1\Longrightarrow  \sigma_{\gamma\gamma}^{excl,\; CDF}<410\;{\mathrm fb},\\
&& \hspace*{-1cm}E_T>2.5\;{\mathrm GeV},\; |\eta_{\gamma}|<1\nonumber\\
\label{CDF2gam2}&&\hspace*{-0.7cm}\Longrightarrow  \sigma_{\gamma\gamma}^{excl,\; CDF}=2.48 {{+0.40}\atop{-0.35}}\;{\mathrm (stat)}\;{{+0.40}\atop{-0.51}}\;{\mathrm (syst)}\;{\mathrm pb}
\end{eqnarray}
are in agreement with these predictions.

 Our prediction for the $\chi_{c,0}$ production is
\begin{equation}
\label{eq:chic3P} \left.\frac{d\sigma^{excl,\; th}_{\chi_{c,0}}}{dy}\right|_{y=0}=15.9\pm 4.1\;{\mathrm nb}
\end{equation}
and the CDF one~\cite{CDFhq} is
\begin{equation}
\label{eq:chicCDF} \left.\frac{d\sigma^{excl,\; CDF}_{\chi_{c,0}}}{dy}\right|_{y=0}=76\pm 10\;{\mathrm (stat)}\;\pm 10\;{\mathrm (syst)}\;{\mathrm nb}.
\end{equation}
Theoretical prediction is about $3\div 5$ times lower. As was mentioned above, the nonperturbative Pomeron-Pomeron and Reggeon-Reggeon fusion should be taken into account since
this is the case of an intermediate invariant mass. For example, it was shown in~\cite{KMRhq1},\cite{KMRhq2}, that the nonperturbative contribution can be of the same order as the
perturbative one. For more exact estimations we have to do similar calculations in our model.

Our approach can be checked once more in the CDF process of exclusive $J/\Psi$ production $p+\bar{p}\to p+J/\Psi+\bar{p}$ which is governed by the 
photon-Pomeron fusion as in EVMP. We can obtain the cross-section as
\begin{eqnarray}
 \left.\frac{d\sigma^{excl,\; th}_{J/\Psi}}{dy}\right|_{y=0}&=&{\cal C}_{CDF}\times \sigma_{\gamma+p\to J/\Psi+p}(W_0)\nonumber\\
 &=&3.51\pm 0.45\;{\mathrm nb},\label{eq:JPsi3P}
\end{eqnarray}
where 
$$
W_0=\sqrt{m_{J/\Psi}\sqrt{s_{CDF}}}\simeq 78\;{\mathrm GeV},\; {\cal C}_{CDF}\simeq 5.3\times 10^{-5}
$$
(see, for example, Ref.~\cite{thCDFJPsi} for details of calculations).
The CDF result~\cite{CDFhq} is
\begin{eqnarray}
 &&\hspace*{-1.7cm}\left.\frac{d\sigma^{excl,\; CDF}_{J/\Psi}}{dy}\right|_{y=0}\nonumber\\
 &&\hspace*{-0.7cm}=3.92\pm 0.25\;{\mathrm (stat)}\;\pm 0.52\;{\mathrm (syst)}\;{\mathrm nb},\label{eq:JPsiCDF}
\end{eqnarray}
which is in a good agreement with the prediction. The prediction~(\ref{eq:JPsi3P}) does not depend much on the model (since the colliding energy $W$
lies in the HERA interval), that is why it is shown only as an illustration.

 Another ``standard candle'' is the di-hadron production. We have not considered this process yet, but one 
 can find some results in~\cite{KMRhh},\cite{LHChh}.
 
\subsection{Standard candles for low invariant masses.}

\begin{figure}[hbt] 
  \includegraphics[width=0.49\textwidth]{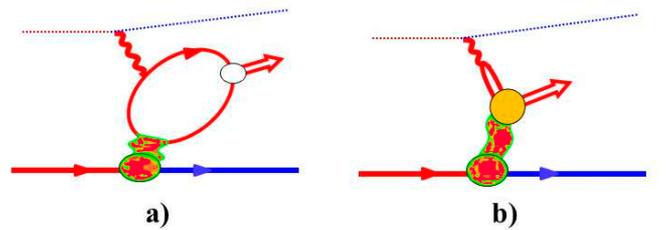} 
  \caption{\label{fig:EVMPother} Representations for EVMP processes with intermediate (a) $M\sim 3-10$~GeV) and low (b) $M\sim 1$~GeV) mass of vector meson.}
\end{figure}

It is shown in the previous section that for invariant masses about several GeV (intermediate case) our model gives results 
systematically lower than the ex\-pe\-ri\-mental data. It is naturally to assume that we have to take into account 
nonperturbative contributions to the Pomeron-Pomeron fusion. We can use, for example, NRQCD as in the case of EVMP, or the method
considered in~\cite{KMRhq2}, where authors represent the Pomeron-Pomeron fusion in a similar way as the gamma-gamma fusion with 
different coupling (see Fig.~\ref{fig:EDDEdyn}b). We could use a similar representation in the EVMP process as depicted in Fig.~\ref{fig:EVMPother}a
to obtain parameters of the model.

The case of $M\sim 1$~GeV is more complicated, because we can not use any (even semi-perturbative like NRQCD) mechanism for
calculations. We can only restrict ourselves by general representations for amplitudes like~(\ref{eq:lowm1})-(\ref{eq:lowm4}) and
fit the data~\cite{WA102}-\cite{WA102d} with these distributions. The production of low mass resonances was considered 
in~\cite{spin_parity_analyser4}. We can use this ``standard candle'' and also EVMP process with light vector mesons depicted in the Fig.~\ref{fig:EVMPother}b to 
obtain parameters of any nonperturbative 
approach. This case is more convenient from the ex\-pe\-ri\-mental point of view, since cross-sections are much larger than
for high invariant masses. This is a powerful tool to
look for new states like ``glueballs'' by the use of the azimuthal 
angular distributions~\cite{spin_parity_analyser3a},\cite{spin_parity_analyser3b}.

\subsection{LHC predictions in the 3-Pomeron model.}

In this section we collect some predictions of the 3-Pomeron model 
for the LHC.

 The first prediction is devoted to the SM Higgs boson production
 \begin{eqnarray}
 &&\hspace*{-0.5cm}\label{eq:LHCHiggs}\sigma_{p+p\to p+H+p}(M_H=125\; {\mathrm GeV})\simeq 0.55\pm 0.15\; {\mathrm fb},\\
 &&\hspace*{-0.5cm} \label{eq:LHCcutsXiT} 10^{-4}<\xi_{1,2}<0.1,\; 0.001\;{\mathrm GeV}^2<|t_{1,2}|<1\;{\mathrm GeV}^2.
 \end{eqnarray}
 This prediction is based on the EVMP HERA data, but the uncertainty in normalization on ``standard candles'' can be larger and reach a factor 
 like ${{\times 3}\atop{\div 3}}$~\cite{KMRHiggs}.
 
\begin{figure}[b!] 
  \includegraphics[width=0.49\textwidth]{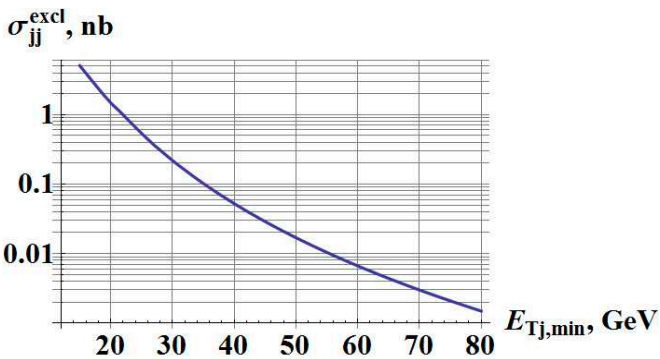} 
  \caption{\label{fig:LHCjj} EDDE di-jet production rate at LHC versus lower cut on the transverse energy of a jet for $\sqrt{s}=8$~TeV, $|\eta_{jet}|<2.5$ plus cuts~(\ref{eq:LHCcutsXiT}).}
\end{figure}

\begin{figure}[hbt] 
  \includegraphics[width=0.49\textwidth]{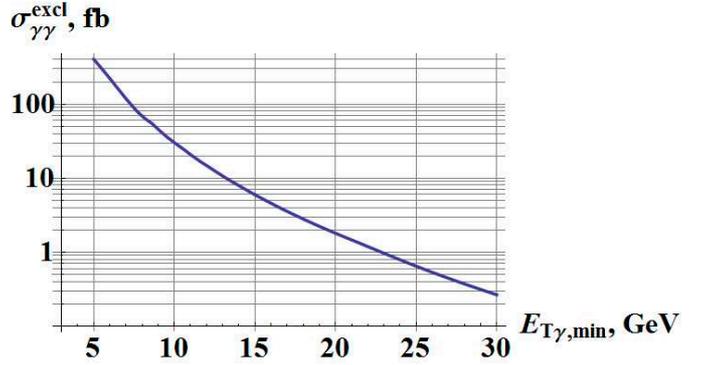} 
  \caption{\label{fig:LHCgamgam} EDDE di-gamma production rate at LHC versus lower cut on the transverse energy of a photon for $\sqrt{s}=8$~TeV, $|\eta_{\gamma}|<2.5$ plus cuts~(\ref{eq:LHCcutsXiT}).}
\end{figure}

 Then we can calculate also di-jet and di-gamma production. Results are presented in
 Figs.~\ref{fig:LHCjj},\ref{fig:LHCgamgam}. For the case of di-gamma production we can assume higher rate (as in CDF) due to nonperturbative 
 effects. The latest LHC results~\cite{LHC2gamCMS} give the upper bound for the EDDE di-gamma production at $\sqrt{s}=7$~TeV
 \begin{eqnarray}
 &&\sigma_{p+p\to p^*+\gamma\gamma+p^*}<1.18\;{\mathrm pb},\;\nonumber\\
 && E_{T,\gamma}>5.5\;{\mathrm GeV},\; |\eta_{\gamma}|<2.5,
\end{eqnarray}
and no particles in the region $|\eta_{\gamma}|<5.2$.

 Now there are preliminary results from LHCb collaboration on exclusive $\chi_{c,0}$ and $J/\Psi$ production at $7$~TeV~\cite{LHCchicLHCb}:
 \begin{eqnarray}
 && \label{eq:chicLHCb} \sigma^{excl,\; LHCb}_{\chi_{c,0}}=160.9\pm 78.8\;{\mathrm nb},\\
 && \label{eq:JPsiLHCb} \sigma^{excl,\; LHCb}_{J/\Psi}=81.9\pm 18.3\;{\mathrm nb},
 \end{eqnarray}
 
 Our predictions are
 \begin{eqnarray}
  \label{eq:chic3PLHC} \left.\frac{d\sigma^{excl,\; th}_{\chi_{c,0}}}{dy}\right|_{y=0}&=& 20\pm 5\;{\mathrm nb},\\
 \label{eq:chic3PLHCtot} 
 \sigma^{excl,\; th}_{\chi_{c,0}}&=& 212\pm 53\;{\mathrm nb},\\
  \left.\frac{d\sigma^{excl,\; th}_{J/\Psi}}{dy}\right|_{y=0}&=& {\cal C}_{LHC}\times \sigma_{\gamma+p\to J/\Psi+p}(W_0)\nonumber\\
 \label{eq:JPsi3PLHC}&=&7.06\pm 0.91\;{\mathrm nb},\\
 \label{eq:JPsi3PLHCtot} 
 \sigma^{excl,\; th}_{J/\Psi}&=& 76.3\pm 19.1\;{\mathrm nb},
  \end{eqnarray}
 where 
 $$
 W_0=\sqrt{m_{J/\Psi}\sqrt{7000\;{\mathrm GeV}}}\simeq 147\;{\mathrm GeV},\; {\cal C}_{LHC}\simeq 6.6\times 10^{-5}.
 $$ 
 These results are close to the ex\-pe\-ri\-mental data.

Results for other interesting processes can be obtained by the use of master formulae~(\ref{eq:EDDEcs}),(\ref{eq:EDDEcsRes}).  

\section{Conclusions and discussions.}

In this article we consider the general framework for the exclusive double diffraction. In the present state
of investigations we can describe it by the use of the prescription depicted in Fig.~\ref{fig:EDDEdyn}:
\begin{itemize}
\item for high invariant masses ($M\gg 1$~GeV) we have to use a perturbative approach for the ``bare'' amplitude and calculate the
gluon loop integral of the product of diffractive proton-gluon amplitudes $T^D$, gluon-gluon fusion vertex and Sudakov-like
suppression factor\linebreak (Fig.~\ref{fig:EDDEdyn}a);
\item for intermediate invariant masses ($M\sim 3\to 10$~GeV like in $\chi_{c}$ production) we can apply the NRQCD approach to the $Q\bar{Q}$-quarkonium vertex and
the gamma-like approximation for a Pomeron~\cite{KMRhq2};
\item for low invariant masses ($M\sim 1$~GeV) we have only the general representation~(\ref{eq:lowm1})-(\ref{eq:lowm4}) for the Pomeron-Pomeron fusion vertex. We can apply also some nonperturbative
approaches like~\cite{spin_parity_analyser3a}-\cite{lowmassmesonsLHC}.
\end{itemize}
After the calculation of the ``bare'' amplitude we have to take into account different rescattering corrections (``soft survival probability'').

Let us note that in this paper we use the old version of 3-Pomeron model only as an example. After
the comparison with the latest TOTEM data~\cite{TOTEM2012a},\cite{TOTEM2012b} it shows discrepancy between
the prediction and data points (as other popular models~\cite{Godizov2012}), and we have to some update 
our approach to better fit the data.

% disadvantages and ``bad'' points
The algorithm for EDDE calculations is far from an ideal one. This is because of our large gaps in understanding of the 
hadronic diffraction. There are different approaches at different scales. Also there are many models for diffractive amplitudes that
are not always adequate (see~\cite{Godizov2012} and references therein). This difficult task needs futher intensive investigations, and
EDDE is the most powerful tool in this scope of activity.

\section*{Appendix A}

Here we collect basic expressions for gluon-gluon fusion partonic cross-sections of the type $g+g\to a+b$. Some of them can be found, for example, in~\cite{ggcs1} ($gg\to gg$, $gg\to Q\bar{Q}$) 
and~\cite{QQth1} ($gg\to\gamma\gamma$).

\begin{eqnarray}
&&\label{HARDggtogg}\frac{d\hat{\sigma}_{gg\to gg}^{J_z=0}}{d\eta}=
\frac{18\pi\alpha_s(M/2)^2\cosh^2\eta}{M^2},\\
&&\label{HARDggtoQQ}\frac{d\hat{\sigma}_{gg\to Q\bar{Q}}^{J_z=0}}{d\eta}=
\frac{4\pi\alpha_s(M/2)^2\cosh^2\eta}{3M^2}\frac{m_Q^2}{M^2}\beta^2,\\
&&\beta=\sqrt{ 1-\frac{4m_Q^2}{M^2}},\nonumber\\
&&\label{HARDggtogamgam}\frac{d\hat{\sigma}_{gg\to \gamma\gamma}^{J_z=0}}{d\eta}=
\frac{121\alpha_e^2\alpha_s(M/2)^2}{324\pi M^2\cosh^2\eta}f_{\gamma\gamma}(\eta),\\
&&f_{\gamma\gamma}(\eta)=1+\left(
1-2\eta\tanh\eta+\frac{\pi^2+4\eta^2}{4}\left( 1+\tanh^2\eta\right)
\right)^2,\nonumber
\end{eqnarray}
where $\eta=(\eta_a-\eta_b)/2$, $M$ is the invariant mass of the system, $\alpha_e$ and $\alpha_s$ are
electromagnetic and strong couplins respectively, $m_Q$ is the quark mass.

We use the following formulae for widths of resonances:
\begin{eqnarray}
&&\Gamma_{H\to gg}=
\frac{M_H^3}{4\pi}
\frac{G_F}{\sqrt{2}}
\left( 
\frac{\alpha_s(M_H/2)}{2\pi}
\right)^2
\left| 
f_H\left( 
\frac{M_H^2}{4m_t^2}\right)
\right|^2K_H,\nonumber\\
&& K_H\simeq 1+\frac{\alpha_s(M_H/2)}{\pi}\left( \pi^2+\frac{11}{2}\right)+0.2,\nonumber\\
&& f_H(x)=\frac{1}{x}\left( 
1+\frac{1}{2}\left( 1-\frac{1}{x}\right)\left[ L_+ +L_-\right]\right),\nonumber\\
&&\label{widthHSM}L_{\pm}=Li_2\left( \frac{2}{1\pm\sqrt{1-\frac{1}{x}}\pm\imath 0}\right),
\end{eqnarray}
\begin{eqnarray}
&&\Gamma_{\chi_b\to gg}\simeq 354.4\;\mathrm{keV} K_{\chi_b},\nonumber\\
&&\label{widthChib}K_{\chi_b}\simeq 
1+\frac{9.8\alpha_s(m_{\chi_b}/2)}{\pi},\; m_{\chi_b}=10.26\;\mathrm{GeV},\\
&&\Gamma_{\chi_c\to gg}\simeq 8.817\;\mathrm{MeV} K_{\chi_c},\nonumber\\
&&\label{widthChic}K_{\chi_c}\simeq 1.69,\; m_{\chi_c}=3.5\;\mathrm{GeV}
\end{eqnarray}
where $G_F$ is the Fermi constant and $m_t$ is the top quark mass.

\section*{Appendix B}

Here we present the calculation of the ``soft survival probability'' using~(\ref{eq:SoftSurv0}) 
in the simple case, when the amplitude has the form
\begin{equation}
\label{survsimple1}{\cal M}\sim \bar{{\cal M}}=\mathrm{e}^{-B(\vec{\Delta}_1^2+\vec{\Delta}_2^2)}.
\end{equation}
When $y=0$ we can take the above form of the amplitude. In this case
\begin{eqnarray}
&&\label{survsimple2}
\int \left| \bar{{\cal M}}\right| d^2\vec{\Delta}\;d^2\vec{\delta}=\frac{\pi^2}{16B^2},\nonumber\\
&&\vec{\Delta}=\frac{\vec{\Delta}_2+\vec{\Delta}_1}{2},\; \vec{\delta}=\frac{\vec{\Delta}_2-\vec{\Delta}_1}{2},
\end{eqnarray}
and
\begin{eqnarray}
 {\cal M}^U\sim \bar{{\cal M}}^U&=&\int\frac{d^2\vec{q}}{(2\pi)^2}\frac{d^2\vec{q}^{\prime}}{(2\pi)^2}d^2\vec{b}\;d^2\vec{b}^{\prime}\mathrm{e}^{\imath\vec{q}\vec{b}+\imath\vec{q}^{\prime}\vec{b}^{\prime}}\nonumber\\
 \label{survsimple3}&\times& \mathrm{e}^{-2B(\vec{\Delta}^2+\vec{\kappa}^2)-\Omega(s,b)-\Omega(s^{\prime},b^{\prime})},
\end{eqnarray}
where $\vec{\kappa}=\vec{q}+\vec{q}^{\prime}+\vec{\delta}$, $b=|\vec{b}|$, $b^{\prime}=|\vec{b}^{\prime}|$,
\begin{eqnarray}
&&\bar{{\cal M}}^U=\mathrm{e}^{-2B\vec{\Delta}^2}
\int\frac{d^2\vec{b}}{2\pi}\mathrm{e}^{-\imath\vec{\delta}\vec{b}-\Omega(s,b)-\Omega(s^{\prime},b)}
\int\frac{d^2\vec{\kappa}}{2\pi}
\mathrm{e}^{\imath\vec{\kappa}\vec{b}-2B\vec{\kappa}^2}\nonumber\\
&& =\frac{\mathrm{e}^{-2B\vec{\Delta}^2}}{4B}
\int\frac{d^2\vec{b}}{2\pi}\mathrm{e}^{-\imath\vec{\delta}\vec{b}-\Omega(s,b)-\Omega(s^{\prime},b)-b^2/(8B)}\nonumber\\
&&\label{survsimple4}=\frac{\mathrm{e}^{-2B\vec{\Delta}^2}}{4B}h(\delta),
\end{eqnarray}
where the function $h$ is presented in~(\ref{eq:EDDESurv3Ph}). Finally we have
\begin{equation}
\label{survsimple5}\int \left| \bar{{\cal M}}^U\right| d^2\vec{\Delta}\;d^2\vec{\delta}=
\frac{\pi^2}{16B^2}\frac{1}{4B}\int\limits_0^{\infty}\left| h(\delta) \right|^2 d\delta^2,
\end{equation}
and 
\begin{equation}
<S^2>\simeq <S^2>_{y=0}=\frac{\int \left| \bar{{\cal M}}^U\right| d^2\vec{\Delta}\;d^2\vec{\delta}}{\int \left| \bar{{\cal M}}\right| d^2\vec{\Delta}\;d^2\vec{\delta}},
\end{equation}
which leads to the expression~(\ref{eq:EDDESurv3P}). The accuracy of this approximation is about 1\%.

\section*{Aknowledgements}

Author thanks to V.~A.~Petrov, A.~V.~Prokudin, A.~A.~Godizov for useful discussions.

%
% BibTeX users please use
% \bibliographystyle{}
% \bibliography{}

\end{document}